\documentclass{article}
\usepackage{amsmath}

\usepackage[utf8]{inputenc}
\usepackage{gensymb}
\usepackage{graphicx}
\usepackage{dblfloatfix}
\usepackage{authblk}

\usepackage{geometry}
\usepackage{color}
\usepackage{xspace}

\usepackage{tikz}
\usetikzlibrary{tikzmark}
\usepackage{amsfonts}
\usepackage{booktabs}
\usepackage{siunitx}
\usepackage[style=nature, citestyle=numeric-comp,sorting=none]{biblatex}
\usepackage{authblk}

\graphicspath{ {./Figures_Version2/} }

\newcommand{\tas}{TaS$_{2}$\xspace}
\newcommand{\nbse}{NbSe$_{2}$\xspace}
\newcommand{\nbs}{NbS$_{2}$\xspace}
\newcommand{\tase}{TaSe$_{2}$\xspace}
\newcommand{\mos}{MoS$_{2}$\xspace}
\newcommand{\sio}{SiO$_{2}$\xspace}
\newcommand{\Hpa}{$H_{||}$\xspace}
\newcommand{\Hpe}{$H_\perp$\xspace}
\newcommand{\Hc}{$H_{C2}$\xspace}

\title{The transition-metal-dichalcogenide family as a superconductor tuned by charge density wave strength}

\author[1,3]{Shahar Simon} 
\author[2]{Hennadii Yerzhakov}
\author[4]{Sajilesh K. P.}
\author[3]{Atzmon Vakahi}
\author[3]{Sergei Remennik}
\author[2]{Jonathan Ruhman}
\author[1]{Maxim Khodas}
\author[1,3]{Oded Millo}
\author[1,3]{Hadar Steinberg}

\affil[1]{\textit{The Racah Institute of Physics, The Hebrew University of Jerusalem, Jerusalem 91904, Israel}}
\affil[2]{\textit{Department of Physics, Bar Ilan University, Ramat Gan 5290002, Israel}}
\affil[3]{\textit{The Center for Nanoscience and Nanotechnology, Hebrew University, Jerusalem 91904, Israel}}
\affil[4]{\textit{Physics Department, Technion -- Israel Institute of Technology, Haifa 32000, Israel}}

\begin{document}

\maketitle

\begin{abstract}
\textbf{Metallic transition metal dichalcogenides (TMDs), consisting of H-\nbse, H-\nbs, H-\tase and H-\tas, remain superconducting down to a thickness of a single layer.
In these materials, thickness affects a variety of properties -- including Ising protection, two-band superconductivity, 
and the critical temperature $T_C$, which decreases for the Nb-based, and increases for the Ta-based materials.
This contradicting trend is puzzling, and has precluded the development of a unified theory.
We approach the question of thickness-evolution of $T_C$ and the superconducting gap $\Delta$ by measuring high-resolution tunneling spectra in \tas-based stacked devices. Our measurements allow for simultaneous evaluation of $\Delta$, $T_C$, and the upper critical field \Hc. 
The latter, we find, is strongly enhanced towards the single-layer limit -- following a $H_{C2} \propto \Delta^2$ proportionality ratio.
Our main finding is that the same ratio holds for the entire family of metallic TMDs: \tas and \nbse of all thicknesses, bulk \tase and bulk \nbs, extending over 4 orders of magnitude in \Hc and covering both clean and dirty limits.
We propose that this tunability across the TMD family is controlled by the competing charge density wave (CDW) phase.
Using Gor'kov's theory, we calculate how a CDW order affects the quasiparticle density of states and the resulting $T_C$ and \Hc. Our results suggest that CDW is the key determinant factor limiting $T_C$ in the TMD family. 
They also show that \Hc is universally enhanced by a factor of two orders of magnitude above the expected value, an effect that remains an open question.
}
\end{abstract} 

\section*{Introduction}
Superconducting transition metal dichalcogenides (TMDs) combine intricate effects of thickness with the physics of superconductivity, owing to the ability to accurately control the sample thickness in a clean manner via exfoliation. 
\nbse, \tas, \nbs and \tase all share a similar band-structure and a 2H crystal structure in their bulk form, and exhibit hole pockets in their transition metal bands around the $K$ and $\Gamma$ points \cite{lian2023four_material_bandstructure,castroneto2001four_materials_cdw_SC}. 
Among these, the most extensively studied materials are \nbse and \tas, where strong spin-orbit coupling (SOC) gives rise to an Ising protection at the ultrathin limit \cite{Xi_KFMak_NbSe2_2016, Hunt_NbSe2_TaS2} and to possible development of a triplet order emerging at high in-plane magnetic fields \cite{Hamill2021, kuzmanovic2022tunneling,mockli2019triplets}.
The symmetries and layered structure also enable exotic inter-layer effects such as the orbital Fulde-Ferrell-Larkin-Ovchinnikov (FFLO) state \cite{wan2023orbital_FFLO_ising,cho2023FFLO_bulk_NbSe2,zhao2023finite_momentum}.

Significant attention was given to the effect of thickness on the superconducting properties of these materials \cite{Xi_KFMak_NbSe2_2016, Tom_spectroscopy_2018, Khestanova2018, Hunt_NbSe2_TaS2,yan2019thickness_nbs2,wu2018thickness_tase2}. 
Specifically, \nbse exhibits a reduction of the critical temperature $T_C$ from 7.2 K at the bulk to 3 K at the monolayer limit~\cite{Xi_KFMak_NbSe2_2016}, accompanied by a suppression of the gap~\cite{Tom_spectroscopy_2018, Khestanova2018}. A similar suppression of $T_C$ is seen in \nbs.~\cite{yan2019thickness_nbs2}. 
Xi et al.~\cite{Xi_KFMak_NbSe2_2016} suggested that this suppression is the result of fewer adjacent layers available to assist in Cooper pairing via interlayer interaction, an effect observed in high-$T_C$ superconductors experimentally \cite{li1990interlayer_coupling_experiment} and also treated theoretically \cite{schneider1991interlayer_coupling_theory}.
Remarkably, in \tas the inverse effect appears, with $T_C$ increasing from 0.8 K in the bulk to 3 K in a single layer, as seen in transport experiments~\cite{Efren_tas2_2016, 
Yafang_tas2_cdw, Hunt_NbSe2_TaS2}.
A similar enhancement is seen in \tase~\cite{wu2018thickness_tase2}.
Navarro-Moratalla et al. \cite{Efren_tas2_2016} suggested that the strength of the effective coupling constant, accounting for electron-phonon coupling and Coulomb repulsion, could vary with thickness and possibly reverse the typical dependence of $T_C$ on thickness. A different approach was taken by Yang et al. \cite{Yafang_tas2_cdw}, who focused on the role of the charge density wave (CDW) order that can suppress superconductivity by gapping segments of the Fermi surface.
All together, the contrasting behavior between Nb- and Ta-based materials raises an important question: whether thickness-dependent superconductivity in these two groups is governed by the same mechanism. 

In this work, we approach the question of thickness-dependent superconductivity in metallic 2H-TMDs by measuring the tunneling spectra of \tas devices of varying thickness, from bulk down to a monolayer. 
Spectra are measured using stacked all-TMD tunnel devices, where a \mos tunnel barrier is placed on top of the \tas layer. 
Measurements are taken at temperatures down to T = 20-30 mK, allowing us to resolve the smaller gaps of the bulk \tas.
We find that the \tas spectrum exhibits a well-behaved hard gap at all thicknesses, and that $\Delta$ is related to $T_C$ as expected for Bardeen-Cooper-Schrieffer (BCS) superconductivity. 
The thin samples exhibit a gap that survives well beyond our maximally attainable in-plane field of 8.5 T. The stability of this gap is a consequence of Ising spin-orbit protection~\cite{Tom_spectroscopy_2018, kuzmanovic2022tunneling}.

Our main result is found when tracking the thickness-dependence of the out-of-plane upper critical field \Hc, which sharply increases towards the thinner samples. We find that \Hc depends quadratically on $T_C$ and on $\Delta$. 
Surprisingly, by compiling data of all other metallic TMD superconductors, we find that \nbse, \nbs, and \tase share the same quadratic dependence with the very same prefactor.
This observation shows that the metallic TMD family shares a common mechanism driving superconductivity. This further suggests the existence of a single mechanism dictating the exact values of \Hc and $T_C$ across the entire family. 
We present a model where the strength of the CDW interaction drives this effective single-material behavior.

\section*{Results}
\subsection*{Spectra at Zero Magnetic Field and Base Temperature}

\begin{figure}[ht!]
            \centering
            \includegraphics[width=1\columnwidth]{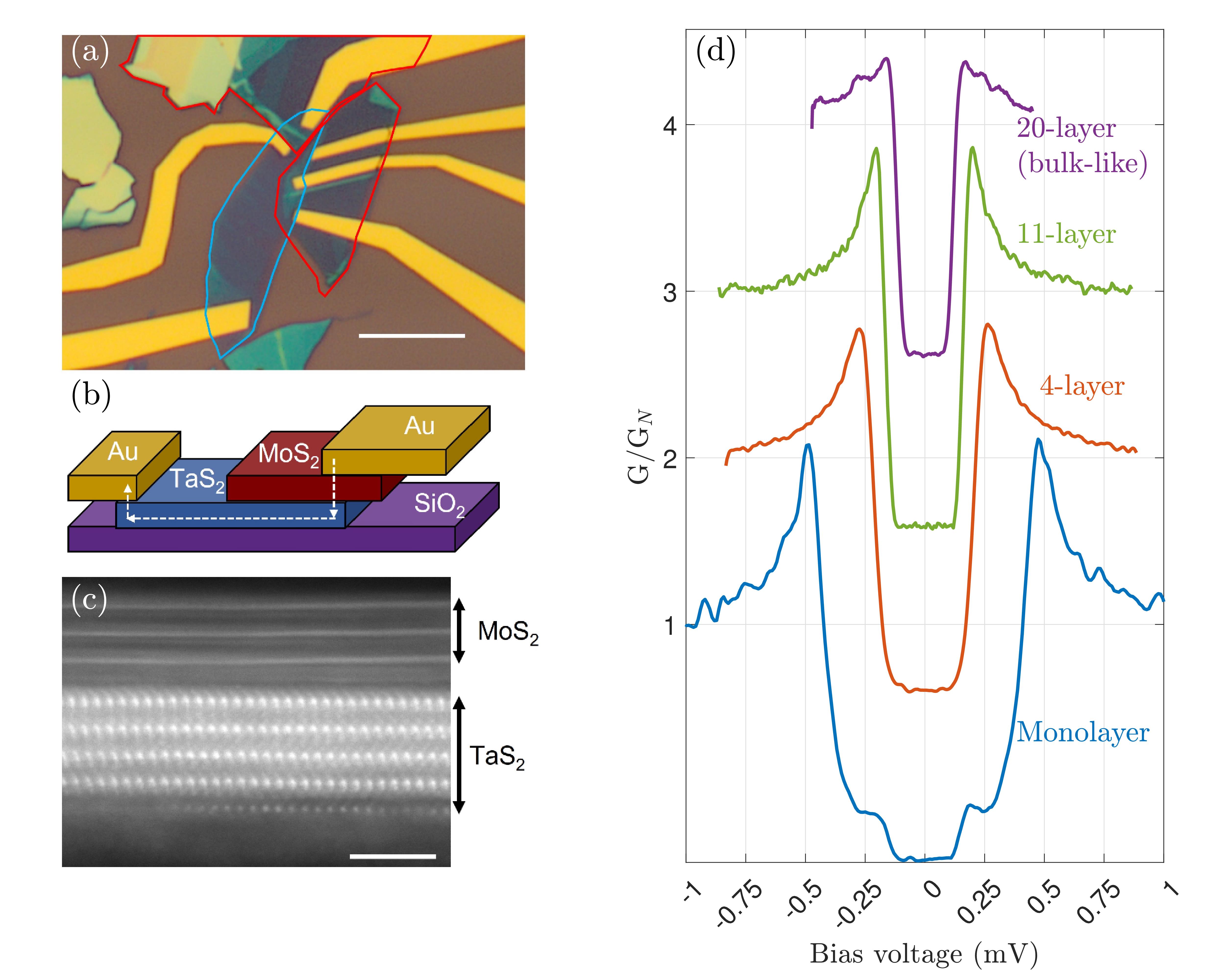}
            \caption{
           (a) Optical microscope image of a typical tunnelling device. The superconductor \tas is outlined in blue and the semiconductor \mos in red. Scale bar: 10 $\mu m$. (b) Simplified schematic of a tunnel junction. \tas is placed on a \sio substrate, and partly covered with a \mos barrier. Ti/Au electrodes are evaporated over the heterostructure (for tunneling contacts) and also on the bare \tas flake (for Ohmic drain contacts). Current path upon application of voltage using the electrodes is indicated by dashed arrows. (c) Cross-section of a 4-layer junction taken in a scanning tunneling electron microscope (STEM). Ta atoms are easily distinguishable as bright dots. A 3-layer \mos barrier is visible on top. Scale bar: 2 $nm$. (d) Representative tunneling spectra at base temperature ($\approx$25 mK) of \tas of different thicknesses. All devices show a hard gap, with diminishing gap width towards thicker samples. 
           }
            \label{fig: Exposition}
        \end{figure}

An optical microscope image of a representative tunneling device is shown in Fig. \ref{fig: Exposition}a, and a corresponding simplified schematic of a junction is presented in panel (b) for clarity, with the current path in white: electrons tunnel from the right Au electrode through the \mos barrier into the \tas flake. 

In order to verify the quality of the interface between materials and accurately determine the number of layers in each junction, we use a focused ion beam (FIB) and a scanning transmission electron microscope (STEM) to image the device cross-section. 
A representative STEM image taken on a 4-layer device is shown in Fig. \ref{fig: Exposition}c. We observe clean, uniform atomic contact over several micrometers of cross-section for all of the junctions that displayed superconducting spectra. 

The tunneling spectra, measured at base temperature ($\approx$ 25 mK) on four tunnel junctions of different \tas thicknesses with $N$=1,4,11,20 ($N$ being the number of layers) are displayed in Fig. \ref{fig: Exposition}d. Additional spectra, measured for all other thicknesses, are displayed in Fig. S1 of the Supplementary Information.
The measurements were carried out using standard lock-in techniques, typically using a 10 $\mu$V  AC voltage excitation. Junctions of all thicknesses exhibit a typical BCS-like tunnel spectrum with its two hallmark features: a strongly suppressed conductance inside the gap (``hard gap"), flanked by sharp quasiparticle peaks.
We note that thin devices exhibit stronger peaks than the thicker ones, but we believe this to be the result of aging in the case of the latter samples.

Throughout this work, one should be careful to distinguish two similar, yet distinct, physical quantities: The first is the spectral gap $\Delta_{spec}$, commonly defined as the distance in energy (or equivalently bias voltage) between the two quasiparticle peaks. The other is the BCS gap $\Delta$, which is the parameter appearing explicitly in BCS theory and is uniquely related to the critical temperature $T_C$ or the electron-phonon interaction strength. These gap parameters coincide at zero temperature and magnetic field. Below, while we directly measure $\Delta_{spec}$, we convert it to $\Delta$ in order to highlight our results' agreement with the BCS model.

We have measured a total of 22 junctions, ranging from a bulk $N$=32 device down to a single layer thick \tas (note that ``layer" refers to a single tri-atomic 1H layer). We observe an increase in $\Delta$ with decreasing $N$, in agreement with the behavior of $T_C$ measured by others in electronic transport \cite{Hunt_NbSe2_TaS2,Efren_tas2_2016}. This trend is in stark contrast with \nbse, in which the largest $\Delta$ and $T_C$ are observed in bulk samples.
The compiled values of $\Delta$ vs. $N$ are presented in Fig. \ref{fig: DeltaTc}b. 
As expected, we measured the largest gap in our monolayer sample,  where  $\Delta \approx 460~\mu eV$, comparable to spectra taken by STM on detached flakes on top of a bulk sample~\cite{galvis2014suderow_detached} and on 4Hb-TaS$_2$ samples, where the 1H-\tas is interleaved with 1T-\tas~\cite{haim_4hb_edge_modes}.
Our results show that thin \tas behaves as a fully gapped BCS superconductor, contrary to the results presented in Ref. \cite{vavno2023liljeroth_f_wave}.
The singular monolayer device measured exhibits a second set of low energy gaps, whose origin is not yet clear to us. These features vanish at temperatures much lower than $T_C$, becoming completely absent at $T = 400~mK = 0.12~T_C$ (see Fig. S2). 
This particular device is further discussed in the Supplementary Information in Section I B. 
Further investigation is required to resolve this issue, which is aside from the main focus of this work.

\subsection*{Thickness Dependence of the Critical Temperature}
\begin{figure}[b!]
            \centering
            \makebox[\textwidth][c]{\includegraphics[width=1.0\textwidth]{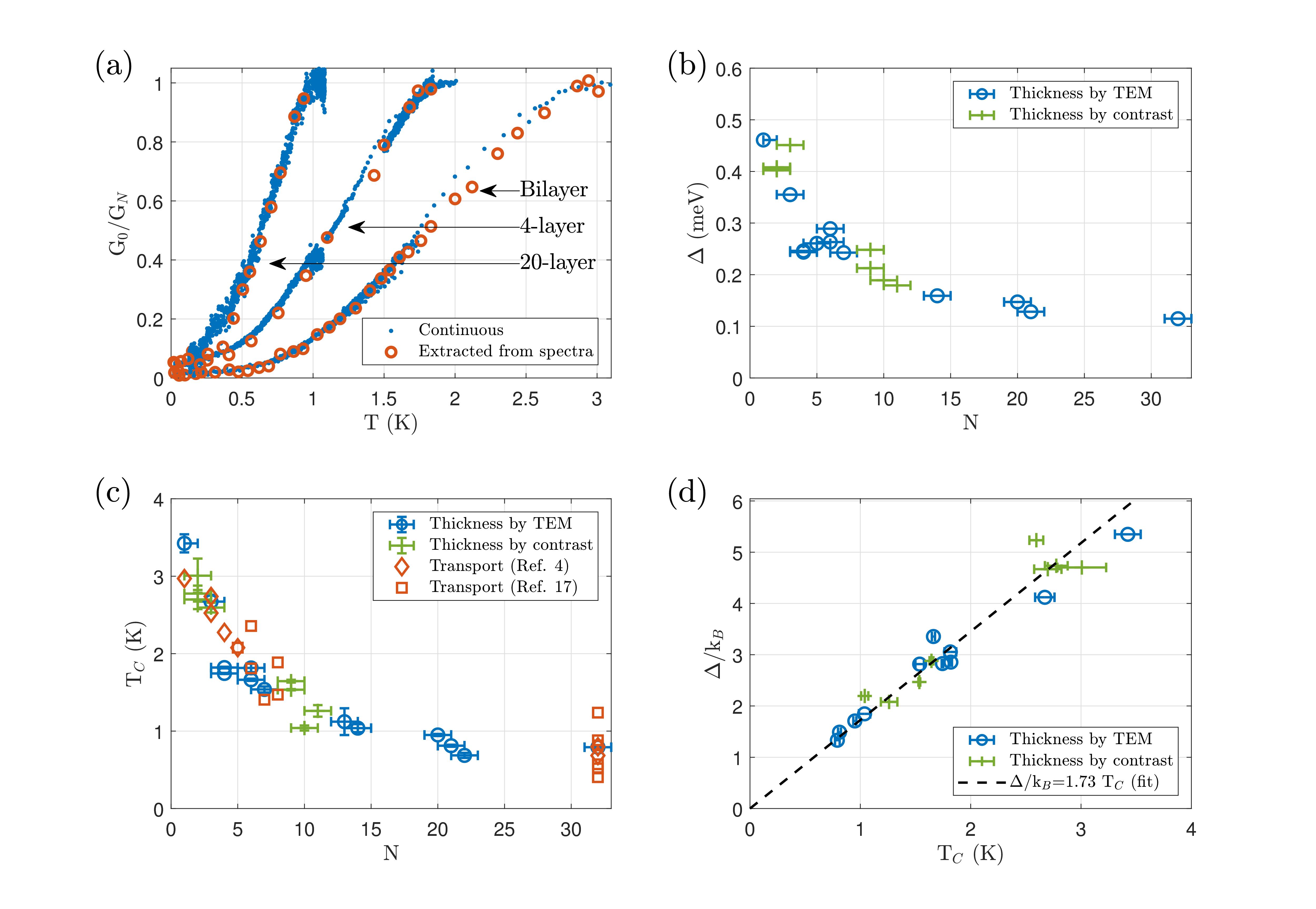}}
            \caption{
            (a) Temperature evolution of the zero-bias tunneling conductance $G_0$, normalized by its value in the normal state $G_N$, of three representative junctions. Plotted in red circles are the zero-bias points extracted from measured spectra. Plotted in blue dots are data measured while continuously cooling the samples from slightly above $T_C$ down to base temperature. (b) BCS gap $\Delta$ vs. $N$. (c) Critical temperature $T_C$ vs. $N$. $T_C$ was taken to be the temperature at which the ratio $G_0/G_N$ reaches 0.95. Overlaid in red are data measured in transport devices by Ref. \cite{Hunt_NbSe2_TaS2} and Ref. \cite{Efren_tas2_2016}, exhibiting good agreement with our data. (d) Dependence of $\Delta/k_B$ on $T_C$. The black line is a linear fit. In all panels, green plus-shaped markers indicate junctions whose thickness could not be directly observed in STEM and had to be estimated based on optical contrast.}
            \label{fig: DeltaTc}
    \end{figure}

Our tunneling measurements allow us to relate $T_C$ to $\Delta$ and compare with BCS theory. 
We measure the evolution of the zero-bias conductance $G_0$ with temperature (see Fig. \ref{fig: DeltaTc}a) and define $T_C$ as the temperature at which $G_0$ reaches $95\%$ of its saturation value $G_N$, giving good agreement with values measured in transport \cite{Hunt_NbSe2_TaS2,Efren_tas2_2016}, see Fig. \ref{fig: DeltaTc}c. 
The dependence of $\Delta$ and $T_C$ on $N$ is shown in Fig. \ref{fig: DeltaTc}b,c. We clearly observe the trend of reduced gap and critical temperature in thicker samples, reaching saturation around $N$ = 20. We then plot the measured $\Delta/k_B$ ($k_B$ is the Boltzmann's constant) vs. $T_C$ in panel (d). A linear fit to these data  gives a slope of 1.73, in good agreement with the weak-coupling BCS model prediction of 1.76.

\subsection*{Tunneling Spectra in Applied Magnetic Field}
We measure the spectra of all devices in the presence of magnetic field, in the plane of the sample (parallel field, denoted \Hpa) or perpendicular to it (denoted \Hpe). 
2H-TMDs are all characterized by a strong Ising SOC~\cite{Xi_KFMak_NbSe2_2016,saito2016, Hunt_NbSe2_TaS2}. Consequently, their electronic spins are locked in an out-of-plane orientation, making superconductivity robust against applied in-plane field. Indeed, similar to thin \nbse, spectra measured in our thinnest samples are only very slightly perturbed by \Hpa. As we show in Fig. S4, the hard gap and quasiparticle peaks persist up to fields exceeding 8.5 T. 
In thicker samples, the finite thickness allows the formation of circulating Meissner currents, whose effect induces pair-breaking~\cite{Tom_spectroscopy_2018}. 
H$_{c||}$ is more easily accessible in thicker samples, and reaches a lower bound of $\approx$ 1.8~T in samples with $N \geq 20$.

\begin{figure}
            \centering
            \makebox[\textwidth][c]{\includegraphics[width=1.0\textwidth]{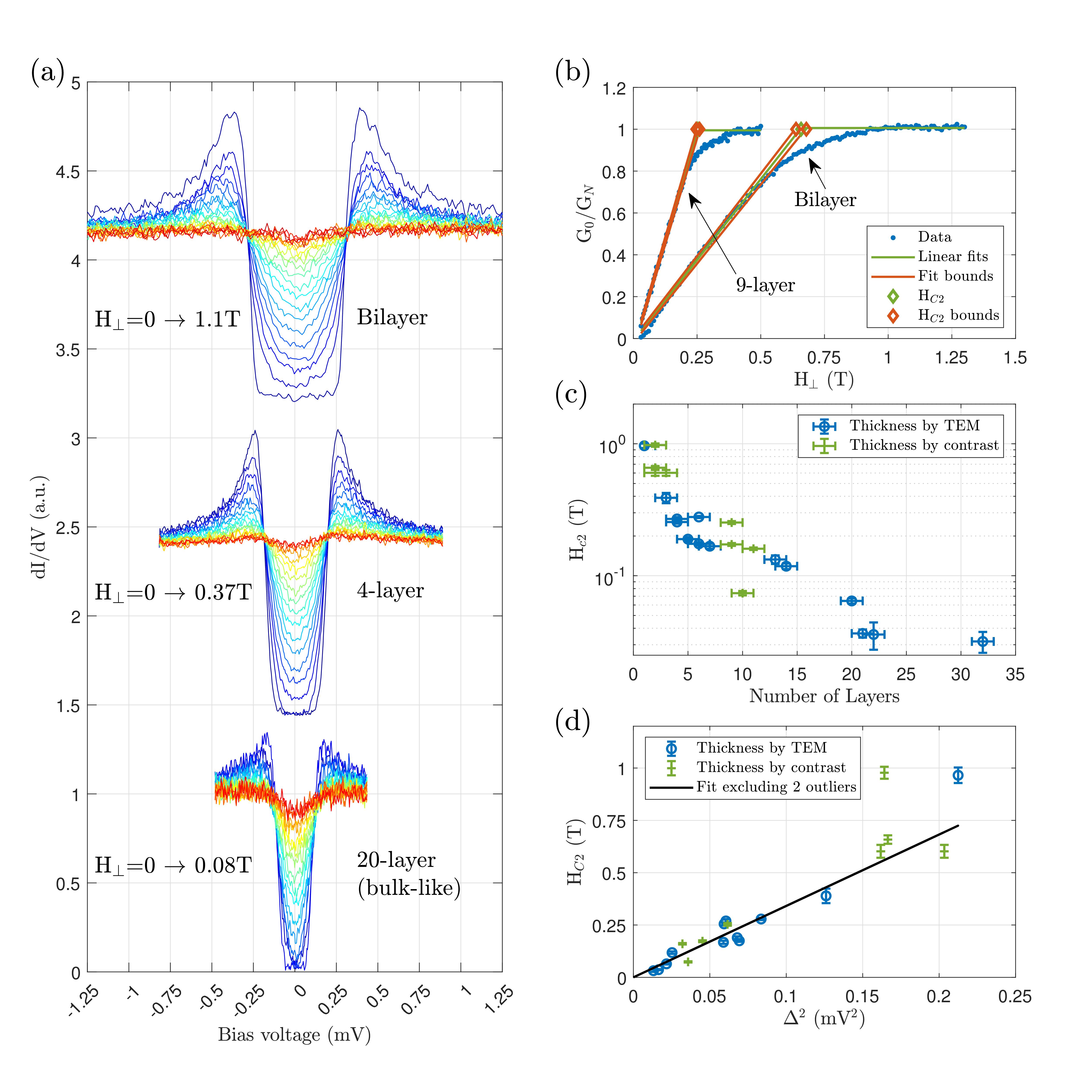}}
            \caption{
           (a) Tunneling spectra of three representative junctions, thickness indicated, in perpendicular magnetic field \Hpe. The spectrum in zero field is plotted in blue and the highest field data in red. 
           (b) $G_0/G_N$ vs. \Hpe for a bilayer and a 9-layer junction. \Hc is defined as the intersection of the two linear regimes. The red diamonds indicate the confidence bounds corresponding to the error bars used in the next panels. (c) \Hc vs. $N$. (d) \Hc vs. $\Delta^2$. 
           }
            \label{fig:HcDelta}
        \end{figure}

We find more striking results when measuring the tunneling spectra in \Hpe. Fig. \ref{fig:HcDelta}a displays the spectra of three representative junctions. The bulk dataset, taken on a 20-layer sample, regains the flat metallic tunneling characteristic by \Hpe = 80 mT. 
In tunneling measurements, where there is no sharp transition to the normal state, we define \Hc as the magnetic field where the two linear regimes of the $G_0$ vs. \Hpe curve intersect (see Fig. \ref{fig:HcDelta}b). 
As we move to thinner samples, \Hc extracted using this method grows to 0.37 T for the 4-layer sample and to 1.1 T for a bilayer, 
in agreement with the values reported in transport measurements~\cite{Hunt_NbSe2_TaS2, Efren_tas2_2016}. 
We note that in some junctions, the high-field spectra retain a small zero-bias dip which remains stable far above saturation. The origin of this dip is not understood, but does not affect our conclusions.
The marked increase in \Hc is seen in panel (c), where we track \Hc vs. $N$.
To glean information about the mechanism leading to such enhancement of \Hc, we plot its value vs. $\Delta^2$ in panel (d). Excluding two outlying points, the dependence appears linear, i.e., \Hc$\propto \Delta^2$.  
As we show below, this is a property expected for clean superconductors, yet it is far more general, and can be followed throughout the entire metallic H-TMD material family -- irrespective of whether the material is considered to be in the clean or dirty limit.

\section*{Discussion}

\begin{figure}[b!]
            \centering
            \makebox[\textwidth][c]{\includegraphics[width=1.0\textwidth]{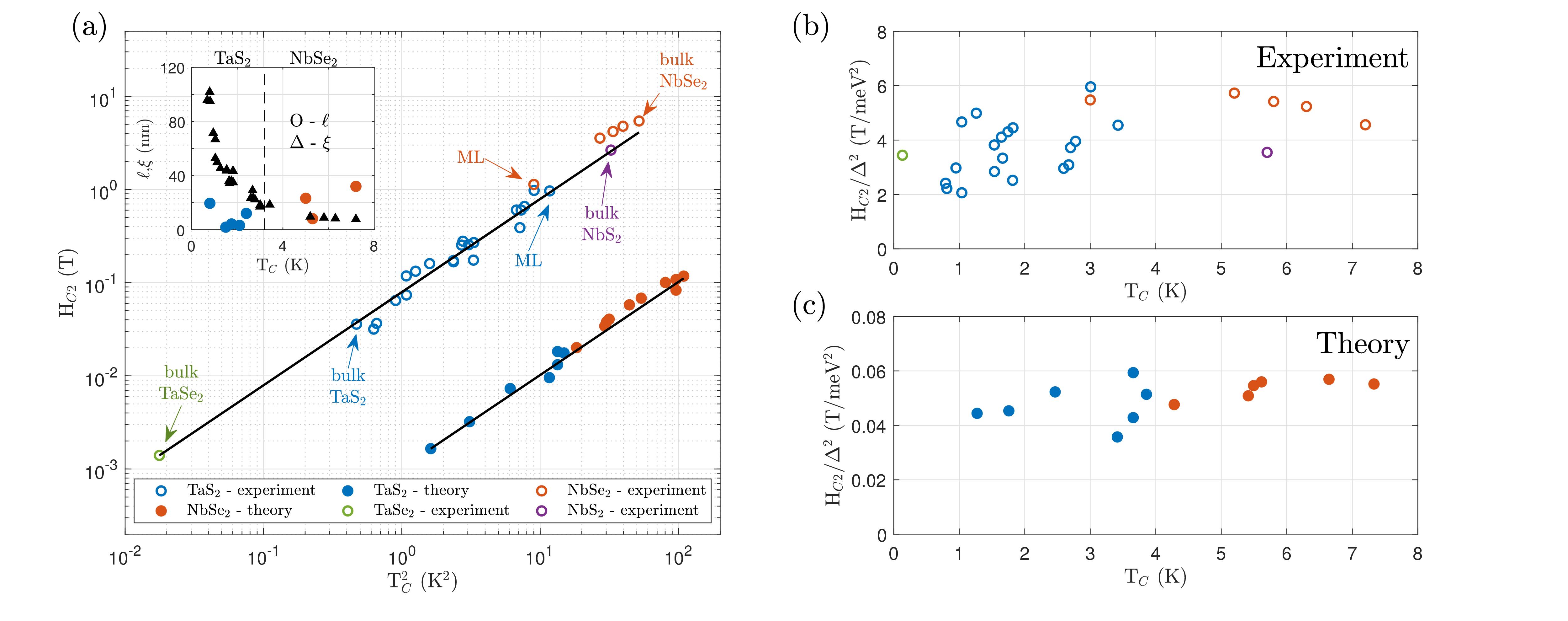}}
            \caption{
           (a) \Hc vs. $T_C^2$ for different TMD superconductors at different thicknesses. Experimental results in empty markers, theoretical results in solid markers. Inset: mean free path $\ell$ (circles, color) and coherence length $\xi$ (triangles, black) for different materials. Blue: \tas, \cite{Efren_tas2_2016,wilson2001tas2_MFP,Hunt_NbSe2_TaS2}; red: \nbse, \cite{Xi_KFMak_NbSe2_2016,Hunt_NbSe2_TaS2,renner1991nbse2_MFP}. 
           (b-c) $H_{C2}/\Delta^2$ vs. $T_C$ ((b) Experiment, (c) Theory, same legend as in (a)).
}
            \label{fig: Theory}
\end{figure}

To understand the thickness dependence of \Hc, we compare \tas to other metallic H-TMD materials. 
We begin by plotting \tas \Hc values vs. $T_C^2$ on a log-log plot in Fig. \ref{fig: Theory}a, noting that $T_C$ is proportional to $\Delta$ (Fig.~\ref{fig: DeltaTc}d). To the TaS$_2$ data points displayed in Fig. \ref{fig:HcDelta}, plotted here in blue, we add the values for NbSe$_2$, plotted in red, which we extract from the transport measurements taken by Xi et al. \cite{Xi_KFMak_NbSe2_2016}. Surprisingly, both data collections appear to arrange on a single common trend-line, reflecting the same $H_{C2}$ vs. $T_C^2$ dependence, with the same pre-factors. 
We note, as seen previously, that whereas in the \nbse data sequence the bulk samples have the higher \Hc and $T_C$, in the \tas sequence the relation is reversed. Curiously, the two materials converge to the same values of $T_C$ and \Hc at the single layer limit.

To these two sequences we also added two more data points:
(i) Bulk TaSe$_2$, where values of $H_{C2}$ = 1.4 mT and $T_C$ = 133 mK were measured by Yokota et al. \cite{yokota_tase2} using the self-inductance method. Remarkably, this single data point resides on the very same line spanned by the TaS$_2$ and NbSe$_2$ sequences. 
(ii) Bulk NbS$_2$, with \Hc = 2.6 T and $T_C$ = 5.7 K as reported in \cite{suderow_nbs2_pressure}. 
Importantly, the solid line superimposed on the data collection marks a quadratic behavior, chosen to cross the TaSe$_2$ data point, and is not a fit. The surprising continuity between sequences belonging to different TMDs of varying thicknesses is the main result of this paper. 
Below, we coin this the ``Metallic H-TMD sequence".

It is then instructive to compare the metallic H-TMD sequence, where we achieve tunability only by varying thickness, with other tuning methods.
In Supplementary Figure S5, we present two additional datasets taken on liquid-gate controlled superconducting MoS$_2$. One \cite{ye2012mos2} exhibits a slope corresponding to $H_{C2} \propto \Delta$, agreeing with a dirty limit superconductor. The other dataset, from Costanzo et al. \cite{costanzo2016} manifests a $\Delta^2$ dependence for the higher $T_C$ range. 
We note that in both these MoS$_2$ studies, the TMDs are electron-doped and are therefore not equivalent to the metallic TMDs, that are hole-doped. 

Unlike gated \mos, in the metallic H-TMD sequence, superconductivity is modulated by both the thickness and the choice of material.
One may conclude that all materials along this sequence share some fundamental traits that govern the nature of their superconductivity, yet are all affected by some `tunable knob' which dictates their position along the sequence.
We suggest that the property, which appears to affect superconductivity so strongly, is the CDW phase that is known to be present to various degrees in all of these materials.

We investigate this possibility by employing Gor'kov's theory to compute \Hc close to  $T_C$, starting from a tight-binding model including SOC and the 3$\times$3 CDW for both \tas and \nbse parameters. 
In this calculation, we assume the attractive phonon-mediated interaction is of equal strength on all three transition metal $d$-orbitals. 
We use the strength of the CDW order parameter, denoted by $b$, as the tuning parameter of $T_C$ and $\Delta$. 
However, it should be noted that the electron-phonon coupling is also expected to vary between different materials. 
We find that $T_C$ decreases monotonically with $b$ due to the reduction of the DOS by the CDW~\cite{flicker2016charge} (for details see Supplementary Information part II).
More importantly, the calculated values of \Hc, shown in Fig.~\ref{fig: Theory}a, are proportional to $T_C^2$, and thus follow the same trend seen in the experimental data up to a significant numerical factor. This discrepancy is explored further below.

We further test the agreement between theory and experiment by plotting the ratio $H_{C2}/\Delta^2$. 
As seen in Fig.~\ref{fig: Theory}b, this ratio varies around a fixed value, a behavior that is also captured by our theoretical model. 
Interestingly, the dependence of the ratio on $T_C$ even resembles that seen in experiment: It increases at small $T_C$ for \tas parameters and decreases for larger $T_C$ with \nbse parameters. 
In contrast, changing the BCS coupling constant by tuning the electron-phonon coupling is not expected to cause such a deviation. 
We thus conclude that the deviation from the sequence implies variations in the band-structure. Indeed, upon tuning the CDW strength $b$, the Fermi velocity $v_F$ varies as well, causing the deviation around the constant value shown in Fig. \ref{fig: Theory}b. 
The accord between theory and experiment highlights the pivotal role of the interplay between CDW and superconductivity in TMD superconductors. 

The comparison with theory also highlights a number of outstanding questions. The observed proportionality relation between \Hc and $\Delta^2$ can be understood if we consider the Ginzburg-Landau (GL) critical field $H_{C2}=\Phi_0/2\pi \xi^2$, where $\Phi_0$ is the flux quantum and $\xi$ the GL coherence length.
In a clean superconductor, one finds $\xi \propto \xi _{BCS} =\hbar v_F/\pi\Delta$, leading to $H_{C2}\propto\Delta^2$ as observed in our data, suggesting that our samples are in the clean limit. 
In this limit we can obtain $v_F$ from the linear fit, and we find $\hbar v_F\approx 0.21~eV\AA$ (corresponding to $v_F\approx3.2\cdot10^4~m/s$), a value that is almost an order of magnitude lower than calculated in density functional theory (DFT) for monolayer \tas or measured in angle-resolved photoemission spectroscopy (ARPES) \cite{lazar2015tas2_DFT,sanders2016tas2_arpes}. This also results in \Hc that is about 80 times higher than expected when taking into account the $v_F$ obtained from the band-structure. This is the discrepancy clearly seen in Fig. \ref{fig: Theory}. It is unclear why disorder does not limit the vortex-core size, leading to a linear dependence of \Hc on $T_C$.

As we show in the inset in Fig. \ref{fig: Theory}, while the values of $\xi$ extracted from the data and the value of the transport mean free path $\ell$ place bulk \nbse in the clean limit \cite{renner1991nbse2_MFP}, in \tas $\ell\approx20~nm$ (extracted using the data in \cite{wilson2001tas2_MFP}), placing it between dirty and intermediate regimes. 
From the comparison between $\ell$ and $\xi$ we conclude this discrepancy is intrinsic and not driven by disorder.  
One possible origin is a velocity renormalization within Fermi liquid theory $v_F^* = v_F/(1+F_1^s/2)$ \cite{pines2018theory}, where $F_1^s$ is the symmetric Landau parameter and $H_{c2}/\Delta^2 \propto  1/(v_F^*)^2$. However, such sizeable discrepancy requires an unphysically large Landau parameter. Another option are strong correlation effects, as seen for example in heavy fermion materials \cite{tachiki1985heavyfermion} or Coulomb repulsion effects \cite{dalal2023CoulombRuhman}. Such an enhancement might reflect the breakdown of weak coupling theory in metallic H-TMDs, e.g. due to spin-fluctuations \cite{das2023Mazin}, although H-TMDs exhibit phenomenological agreement with BCS theory seen in other quantities. The dramatic deviation of metallic H-TMDs from BCS theory therefore poses an outstanding question. 
Our results thus present the H-TMD sequence as a unified highly tunable superconducting system, allowing for testing the properties of superconductivity and its interplay with the CDW phase without any change to carrier density or pressure.

\section*{Acknowledgements}
The authors wish to thank V. Fatemi and A. Kanigel for illuminating discussions. SS is supported by the Milner fellowship for graduate students. 
HS acknowledges support by Israeli Science Foundation grants 861/19 and 164/23.
OM thanks support by ISF (grant no. 576/21) and the Harry de Jur Chair in Applied Science.
JR is supported by the Israeli Science Foundation Grant No.
3467/21.
MK is supported by by the Israeli Science Foundation Grant No. 2665/20.

\printbibliography     
\end{document}


\part*{Supplementary Information}
\tableofcontents
\section{Extended data}

\subsection{Spectra of all devices}
\begin{figure}
            \centering
            \includegraphics[width=1\columnwidth]{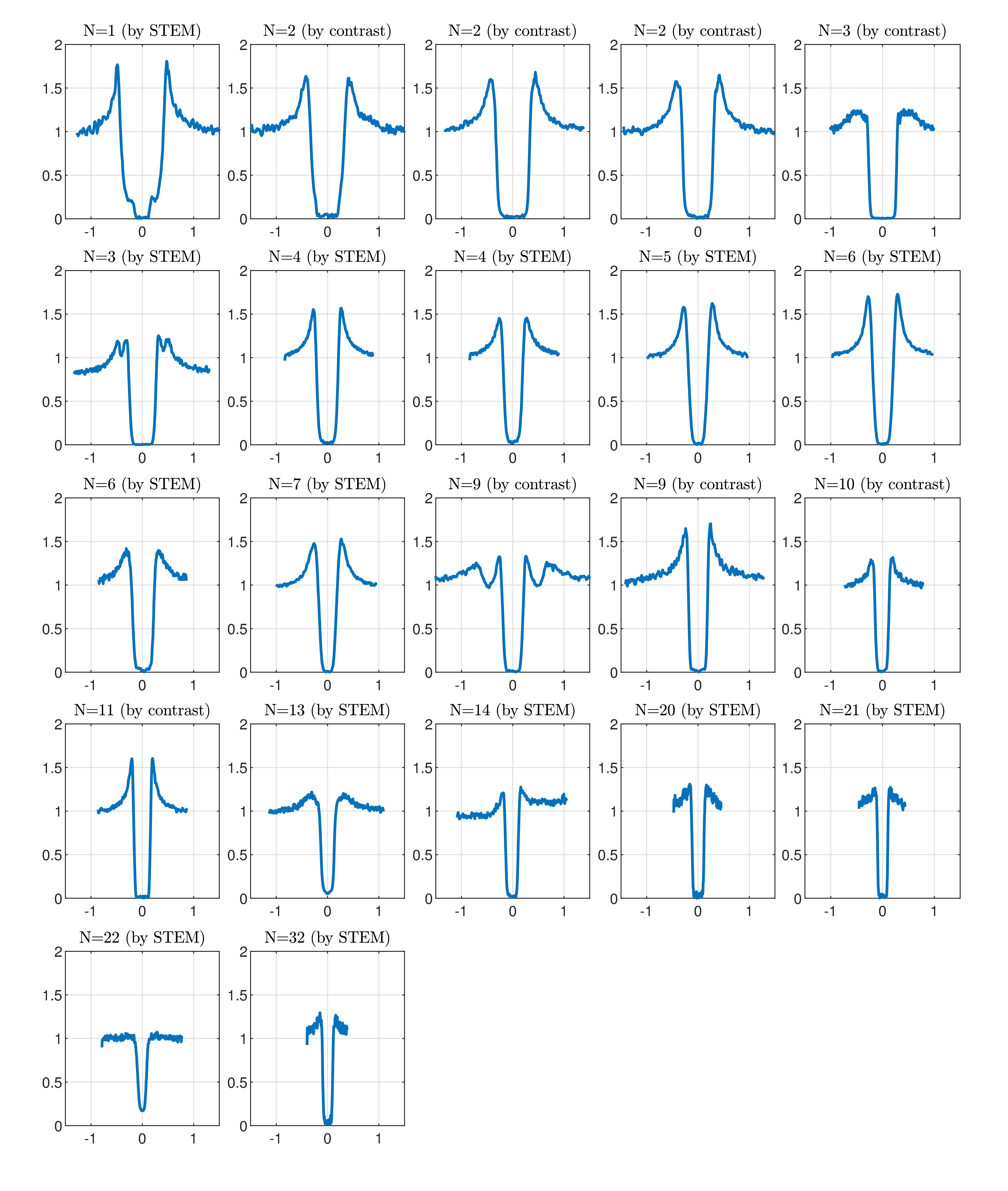}
            \caption{Tunneling spectra of all devices at base temperature and zero field.}
            \label{fig: AllSpectra}
        \end{figure}

Fig. \ref{fig: AllSpectra} shows the base temperature, zero-field tunneling spectra of all \tas devices of thicknesses between one to 32 layers. The horizontal axes are all bias voltage (in units of mV) and the vertical axes are the differential conductance dI/dV, normalized by the value in the normal regime.

\subsection{The in-gap features of the monolayer device}
\begin{figure}
            \centering
            \includegraphics[width=1\columnwidth]{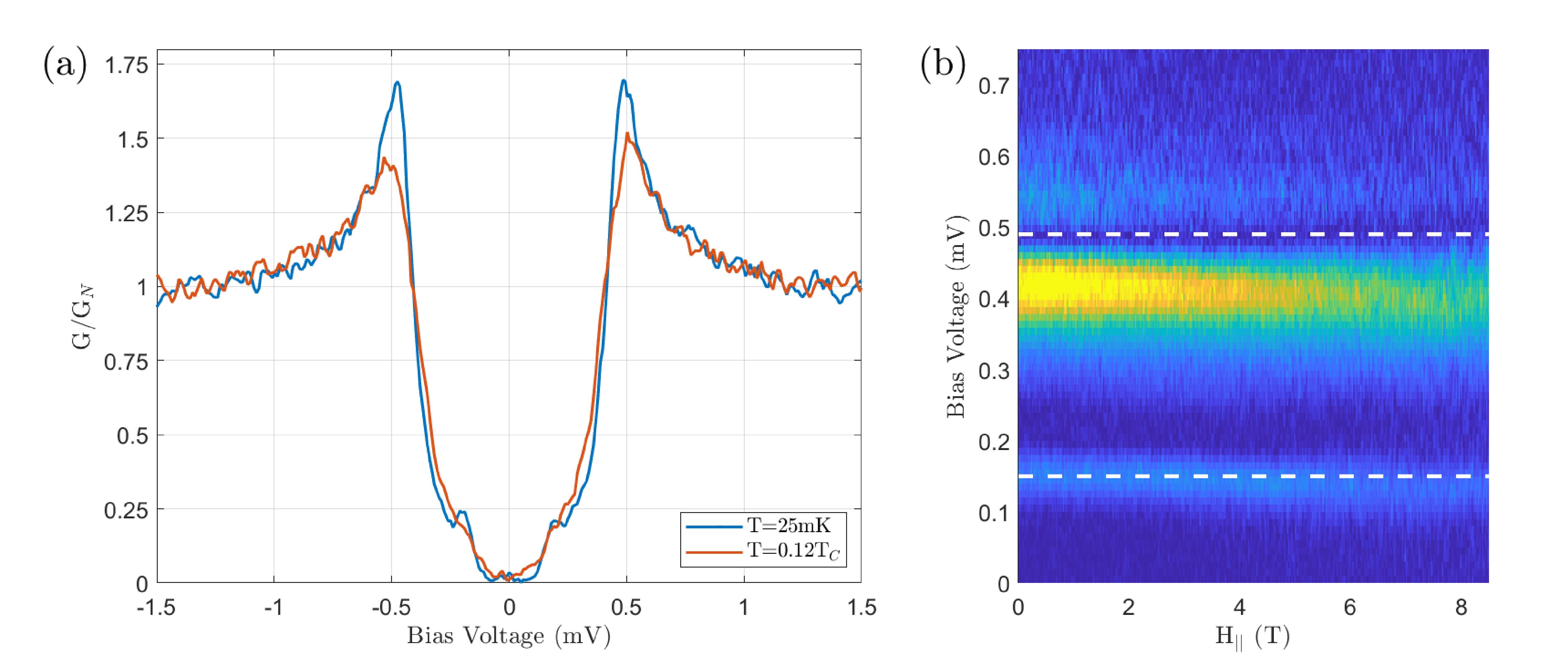}
            \caption{The spectrum of a \tas monolayer junction. (a) The spectrum at base temperature and at T=0.4K, equivalent to $\approx$0.12$T_C$, where the inner feature becomes undetectable. (b) The derivative of the spectrum d$^2$I/dV$^2$ at base temperature,  as a function of in-plane field (only the positive bias is shown for clarity). The dashed lines indicate constant voltage, highlighting the main peak and the inner feature.}
            \label{fig: monolayer}
        \end{figure}

The singular measured monolayer devices exhibits clear in-gap features that disappear at a temperature 0.12$T_C \approx $ 0.4 K (see Fig. \ref{fig: monolayer}a), reminiscent of a smaller, second gap, such as the Se-derived gap observed in \nbse~ \cite{Tom_spectroscopy_2018}. 
However a chalcogen-derived second order parameter cannot be expected in \tas, since the chalcogen band is not populated. 
The identification of the sub-gap feature with a second gap is also not supported by the fact that up to the highest field measured (8.5T), this feature closely follows the behavior of the main gap (Fig. \ref{fig: monolayer}b), in contrast to the effect seen in \nbse where the two gaps respond qualitatively differently to magnetic field due to the distinct dimensionality and different diffusion parameters.

\subsection{Temperature dependence of the spectrum}
\begin{figure}
            \centering
            \includegraphics[width=1\columnwidth]{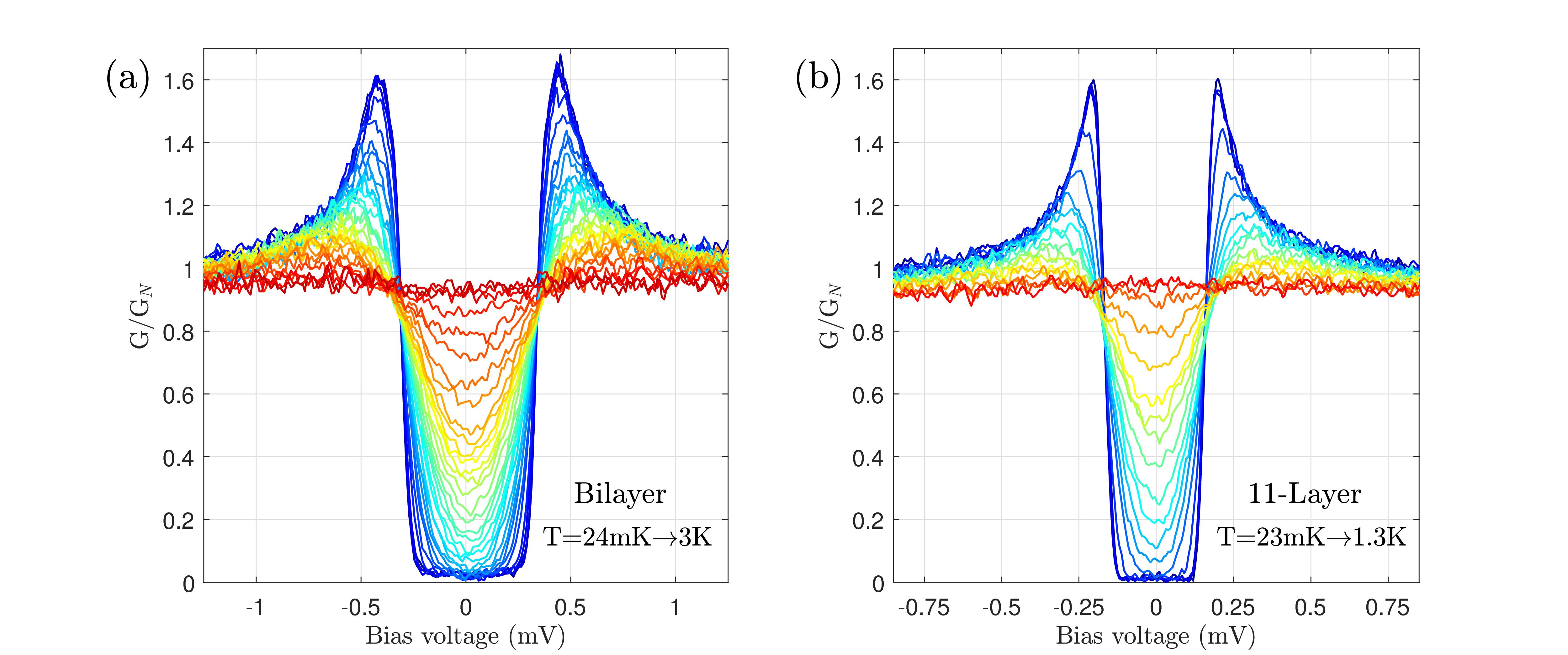}
            \caption{The spectra of (a) a bilayer junction and (b) an 11-layer junction as a function of temperature, with blue (red) indicating lower (higher) temperature. The full range of temperatures is indicated in the panels. 
            }
            \label{fig: supp_tempdep}
        \end{figure}

We measure the spectrum of each junction as a function of temperature, ranging from the cryostat base temperature up to slightly above the temperature at which the spectrum becomes flat, indicating normal metallic behavior. The simultaneous effects of the temperature dependence of $\Delta$ and the smearing of the spectrum due to thermal fluctuations of the order $k_BT$ result in filling of the gap and reduction of the height of the quasiparticle peaks, effectively smearing the spectrum until it becomes metallic. The spectra of two junctions at various temperatures are shown in Fig. \ref{fig: supp_tempdep}. We note a small mismatch between the continuously measured zero-bias conductance $G_0$ during cooldowns and the values extracted from the full spectra (see Fig. 2a of the main text). This mismatch likely originates from the small temperature gradient between the sample and its temperature sensor, expressed in the data measured during cool-down being shifted slightly to colder temperatures compared to those extracted from the slower scans in which we measure a full spectrum at a more stable temperature.

\subsection{Parallel Magnetic Field}
\begin{figure}[p]
            \centering
            \includegraphics[width=1\columnwidth]{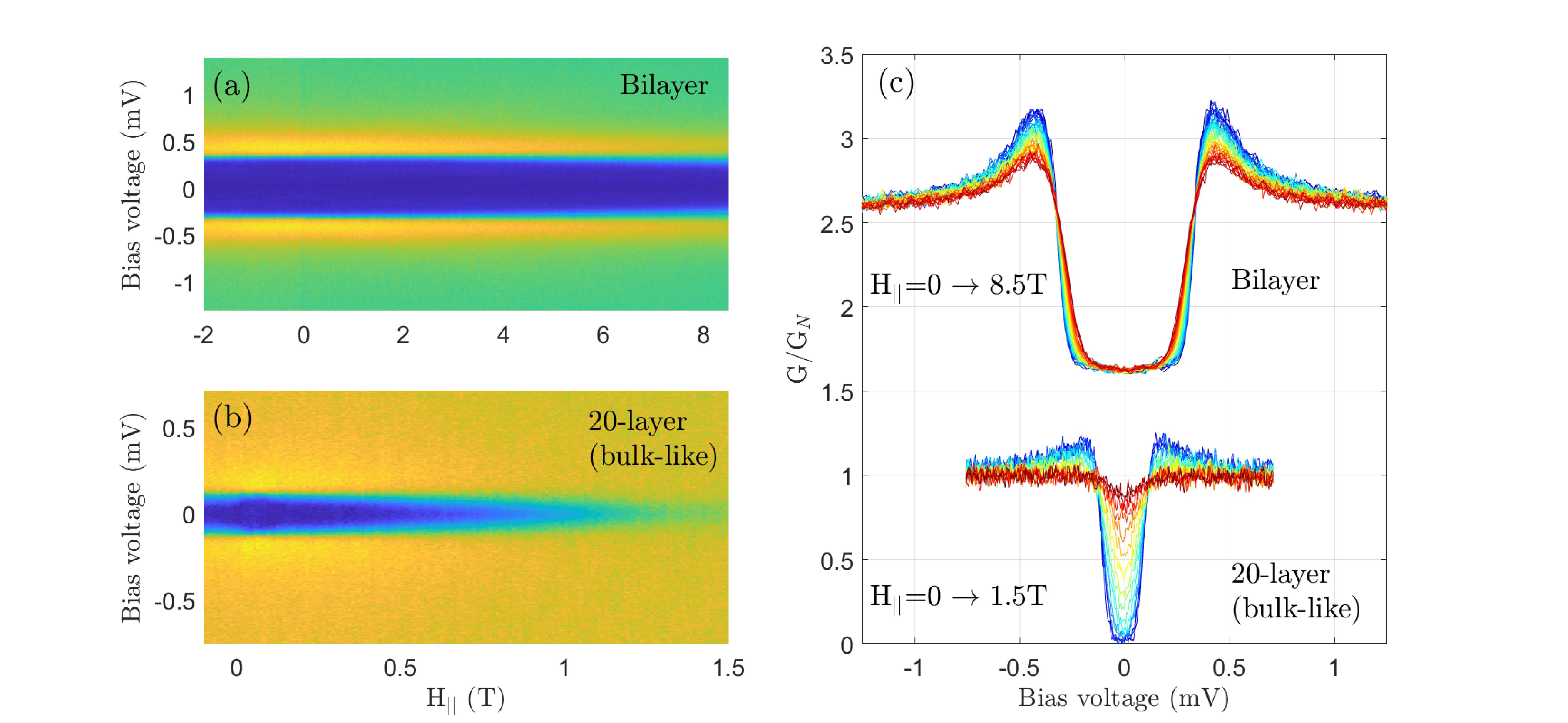}
            \caption{Parallel field response of (a) a bilayer and (b) a 20-layer (bulk) device. Ising SOC is evident in the robustness of the bilayer gap. (c) Overlaid individual spectra from (a,b), with zero field in blue and high fields in red. The full range of fields is indicated in the panels.
            }
            \label{fig: supp_Hpara}
        \end{figure}

Fig. \ref{fig: supp_Hpara} displays the spectra of a bilayer and a 20-layer sample (bulk) under parallel magnetic field. The effect of Ising SOC is evident in the remarkably small effect of fields up to 8.5T on the bilayer sample. The bulk sample, on the other hand, is much more noticeably affected.

\subsection{Metallic H-TMD Sequence with Additional Materials}
\begin{figure}[p]
            \centering
            \includegraphics[width=0.75\columnwidth]{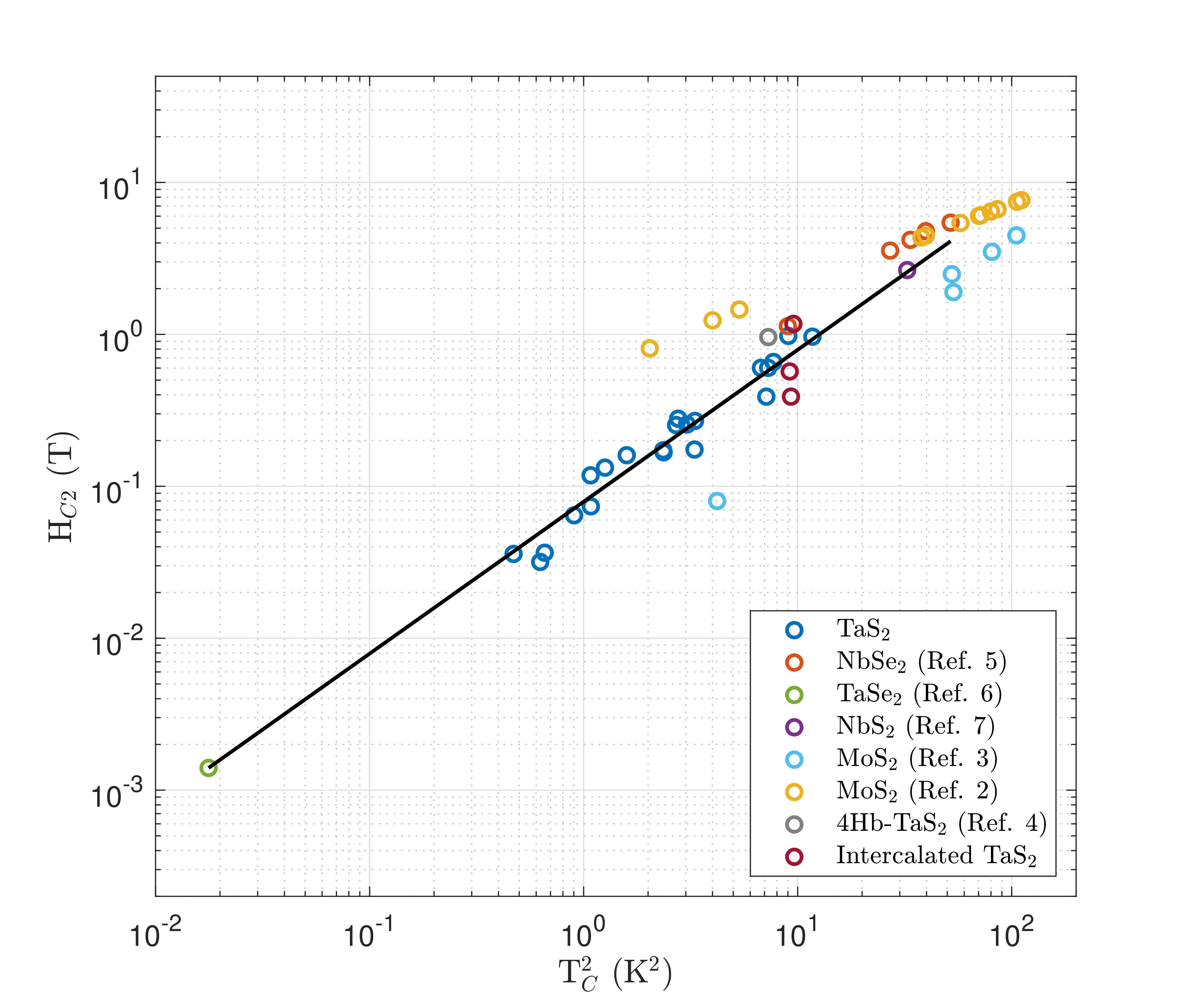}
            \caption{\Hc vs. $T_C^2$ relation of various TMDs. Notably, the metallic H-TMD sequence retains \Hc$ \propto T_C^2$ with the same slope, liquid-ion gated \mos can display either the dirty-limit result \Hc$ \propto T_C$ or the clean-limit result in different experiments (but with a different slope than the metallic ones). Less conventional \tas compounds such as intercalated samples or the 4Hb polytype can behave quite differently. Data adapted from \cite{Xi_KFMak_NbSe2_2016,yokota_tase2,suderow_nbs2_pressure,costanzo2016,ye2012mos2,ribak2020_4hb}.
            }
            \label{fig: metallicHTMDadditional}
        \end{figure}

Fig. \ref{fig: metallicHTMDadditional} displays the \Hc vs. $T_C$ data presented in Fig. 4 of the main text along with additional datasets: 
\begin{itemize}
    \item Liquid-ion gated \mos \cite{ye2012mos2}, displaying \Hc$\propto\Delta$ behavior indicative of dirty-limit superconductivity;
    \item Liquid-ion gated \mos \cite{costanzo2016}, displaying \Hc$\propto\Delta^2$ behavior but with a different constant of proportionality than the metallic H-TMD Sequence;
    \item Data measured on 4Hb-TaS$_2$ \cite{ribak2020_4hb}, which is a qualitatively different superconductor with reports of various exotic phenomena;
    \item Intercalated \tas compounds: \tas intercalated with pyridine molecules (highest $T_C$), and misfit compounds of 1H-\tas sheets separated by square layers of PbS (middle) and SnS (lowest). 
\end{itemize}
        
\section{Theoretical Computation of the upper critical field}
\subsection{Tight-binding model}
\subsubsection{Without CDW}
To model TMD materials in the absence of CDW order, we use the 3-orbital tight-binding model developed in Ref.~\cite{LiuThreeBandTBModel}.  In the basis ${\ket{\psi_1}}={\ket{d_{z^2}},\ {\ket{\psi_2}}=\ket{d_{xy}},\ {\ket{\psi_3}}=\ket{d_{x^2-y^2}}}$, the normal state tight-binding Hamiltonian is given by

\begin{align}
    H_{\b{k}}=\sigma_0 \otimes H_{0\b{k}}+\frac{\lambda}{2}\sigma_z \otimes L_z, 
\end{align}
where $L_z$ in the spin-orbit-coupling term is the z-component of angular momentum operator
\begin{align}
    L_z=\begin{pmatrix}
        0 & 0 & 0 \\
        0 & 0 & -i \\
        0 & i & 0
    \end{pmatrix},
\end{align}
and
\begin{align}H_{0\b{k}}=\begin{pmatrix} V_0 & V_1 & V_2 \\
    V_1^* & V_{11} & V_{12} \\
    V_2^* & V_{12}^* & V_{22}
    \end{pmatrix},
\end{align}
where
\begin{align} 
 V_0=& \epsilon_1+2 t_0(2 \cos \alpha \cos \beta+\cos 2 \alpha) +2 r_0(2 \cos 3 \alpha \cos \beta+\cos 2 \beta) +2 u_0(2 \cos 2 \alpha \cos 2 \beta+\cos 4 \alpha) \\ 
V_1 =& -2 \sqrt{3} t_2 \sin \alpha \sin \beta+2\left(r_1+r_2\right) \sin 3 \alpha \sin \beta - 2 \sqrt{3} u_2 \sin 2 \alpha \sin 2 \beta \\ 
&+ i \left( 2 t_1 \sin \alpha(2 \cos \alpha+\cos \beta)+2\left(r_1-r_2\right) \sin 3 \alpha \cos \beta +2 u_1 \sin 2 \alpha(2 \cos 2 \alpha+\cos 2 \beta) \right), \\
V_2 =& 2 t_2(\cos 2 \alpha-\cos \alpha \cos \beta) -\frac{2}{\sqrt{3}}\left(r_1+r_2\right)(\cos 3 \alpha \cos \beta-\cos 2 \beta) +2 u_2(\cos 4 \alpha-\cos 2 \alpha \cos 2 \beta) \\ 
& + i \left( \sqrt{3} t_1 \cos \alpha \sin \beta +\frac{2}{\sqrt{3}} \sin \beta\left(r_1-r_2\right)(\cos 3 \alpha+2 \cos \beta) +2 \sqrt{3} u_1 \cos 2 \alpha \sin 2 \beta \right), \nn \\
V_{11}=& \epsilon_2+\left(t_{11}+3 t_{22}\right) \cos \alpha \cos \beta+2 t_{11} \cos 2 \alpha +4 r_{11} \cos 3 \alpha \cos \beta+2\left(r_{11}+\sqrt{3} r_{12}\right) \cos 2 \beta \\ \nn
& +\left(u_{11}+3 u_{22}\right) \cos 2 \alpha \cos 2 \beta+2 u_{11} \cos 4 \alpha, \\
V_{12} =& \sqrt{3}\left(t_{22}-t_{11}\right) \sin \alpha \sin \beta+4 r_{12} \sin 3 \alpha \sin \beta +\sqrt{3}\left(u_{22}-u_{11}\right) \sin 2 \alpha \sin 2 \beta, \\
& + i \left( 4 t_{12} \sin \alpha(\cos \alpha-\cos \beta) +4 u_{12} \sin 2 \alpha(\cos 2 \alpha-\cos 2 \beta) \right), \\ \nn
V_{22}= & \epsilon_2+\left(3 t_{11}+t_{22}\right) \cos \alpha \cos \beta+2 t_{22} \cos 2 \alpha +2 r_{11}(2 \cos 3 \alpha \cos \beta+\cos 2 \beta) +\frac{2}{\sqrt{3}} r_{12}(4 \cos 3 \alpha \cos \beta-\cos 2 \beta) \\
& +\left(3 u_{11}+u_{22}\right) \cos 2 \alpha \cos 2 \beta+2 u_{22} \cos 4 \alpha,
\end{align}
with $\alpha = k_x a/2$ and $\beta = \sqrt{3} k_y a/2$, where $a$ is the lattice constant.

\subsubsection{With the CDW}

According to Ref.~\cite{flicker2016charge}, in the commensurate 3Q CDW state the metal atoms' distortion is 
\begin{align}
    u(\b{R}_a)=\sum_{i=1}^{3} u_0 \hat{q}_i \cos(\b{q}_i \cdot \b{R}_a + \phi)\,,
\end{align}
where $\b{q}_1=\frac{\b{b}_1}{3}$, $\b{q}_2=\frac{-\b{b}_1+\b{b}_2}{3}$, $\b{q}_3=-\frac{\b{b}_2}{3}$ are three CDW-vectors. Here $\b{b_1},\b{b_2}$ are primitive vectors in reciprocal space forming 1BZ of the system in the absence of CDW. In the CDW phase, the little BZ (lBZ) is formed by vectors $\b{b_1}/3$ and $\b{b_2}/3$.

In principle, one can derive the CDW-term completely from  symmetry considerations. But to avoid mistakes and aid ourselves with something tangible, we assume that CDW order develops due to electron-phonon interaction (which is likely true\cite{flicker2016charge}):
\begin{align}
    V_{e-p}=\sum_a V_0(\b{r}-\b{R_a}-\b{u_a})-\sum_a V_0(\b{r}-\b{R_a})=-\sum_a \grad V_0(\b{r}-\b{R_a})\cdot \b{u_a}\,.
\end{align}
Consequently, the parameters of the tight-binding model are renoromalized. This can be accounted for by computing the matrix element for scattering between  $\b{k'}$ and $\b{k}$ due to the CDW potential
\begin{align}
    -\mel{\psi_i(\b{k})}{V_{e-p}}{\psi_j(\b{k'})}&=\sum_a \int d^d \b{r} \psi_{i\b{k}}(\b{r})^* \grad V_0(\b{r}-\b{R_a})\cdot \b{u_a} \psi_{j\b{k'}}(\b{r})  \\ \nn
&=    \sum_a \frac{1}{N^3} \sum_{\b{R},\b{R'}}  e^{i \b{k'} \b{R'} - i \b{k} \b{R} }\int d^d \b{r} \phi_{i}(\b{r})^* \grad V_0(\b{r}-(\b{R_a}-\b{R}))\cdot \b{u_a} \phi_{j}(\b{r}-(\b{R'}-\b{R}))  \\ &=\nn
    \sum_a \frac{1}{N^3} \sum_{\b{R},\b{R'}}  e^{i \b{k'} \b{R'} - i \b{k} \b{R} }\int d^d \b{r} \phi_{i}(\b{r})^* \grad V_0(\b{r}-\b{R_a})\cdot \b{u_a}(\b{R_a}+\b{R}) \phi_{j}(\b{r}-(\b{R'}-\b{R})).
\end{align}
Concentrating on CDW-vector $\b{q} \in \{\b{q_1},\b{q_2},\b{q_3}\}$ such that $\b{k}=\b{k'}+\b{q}$, and denoting $\gamma_a(\b{r},\b{R})=\grad V_0(\b{r}-\b{R_a})\cdot \b{u_a}(\b{R_a}+\b{R})$, $\gamma(\b{r},\b{R})=\sum_a \gamma_a(\b{r},\b{R})$, we obtain

\begin{align}
    -\mel{\psi_i(\b{k'}+\b{q})}{V_{e-p}}{\psi_j(\b{k'})} = \frac{1}{N^3} \sum_{\b{R}}  e^{- i \b{q} \b{R} } \left(\sum_{\b{R'}} e^{i \b{k'} \b{R'}} \int d^d \b{r} \phi_{i}(\b{r})^* \gamma(\b{r},\b{R}) \phi_{j}(\b{r}-\b{R'})) \right).
\end{align}

Finally, denoting $F_n(\b{r})=\sum_{\b{R'}} e^{- i \b{q_n} \b{R'}} \gamma(\b{r},\b{R'})$ and $A^n_{ij}(\b{R}) = \frac{1}{N^3} \int d^d \b{r} \phi_{i}(\b{r})^* F_n(\b{r}) \phi_{j}(\b{r}-\b{R}))$, we obtain

\begin{align}
    \Lambda_{\b{q}_n}=\mel{\psi_i(\b{k}+\b{q_i})}{V_{e-p}}{\psi_j(\b{k})} = - \sum_{\b{R}} e^{i \b{k} \b{R} } A^n_{ij}(\b{R}).
\end{align}

Now, we will try to establish the relation between $\Lambda_{\b{q_i}}$, $i=1,2,3$.

First, denoting
\begin{align}
    f_1^i(\b{r})= \sum_a \grad V_0(\b{r}-\b{R_a})\cdot \hat{q}_i \cos(\b{q}_i \cdot \b{R}_a + \phi), \\ \nn
    f_2^i(\b{r})= \sum_a \grad V_0(\b{r}-\b{R_a})\cdot \hat{q}_i \sin(\b{q}_i \cdot \b{R}_a + \phi),
\end{align}
we can write
\begin{align}
    F_i(\b{r})=u_0 \sum_{\b{R'}} e^{- i \b{q_i} \b{R'}} \sum_j \left( f_1^j(\b{r})\cos(\b{q_j}\cdot \b{R'}) - f_2^j(\b{r})\sin(\b{q_j}\cdot \b{R'})  \right). 
\end{align}
Next, denoting 
\begin{align}
    g_{ij}^1= \sum_{\b{R'}} e^{- i \b{q_i} \b{R'}} \cos(\b{q_j}\cdot \b{R'}), \\ \nn
    g_{ij}^2= \sum_{\b{R'}} e^{- i \b{q_i} \b{R'}} \sin(\b{q_j}\cdot \b{R'}),
\end{align}
we find, noting that $\b{q_1}+\b{q_2}+\b{q_3}=0$,
\begin{align}
    g_{ij}^1+ i g_{ij}^2 & = \sum_{\b{R'}} e^{- i \b{q_i} \b{R'}} e^{i \b{q_j}\cdot \b{R'}} = N^3 \delta_{ij}, \\ \nn
    g_{ij}^1- i g_{ij}^2 & = \sum_{\b{R'}} e^{- i \b{q_i} \b{R'}} e^{-i \b{q_j}\cdot \b{R'}} =0.
\end{align}
From which we obtain
\begin{align}
    g_{ij}^1=\frac{N^3}{2} \delta_{ij},\ \  g_{ij}^2=-i \frac{N^3}{2} \delta_{ij}.
\end{align}
Thus, the expression for $F_i(\b{r})$ simplifies to
\begin{align}
\label{eq: Fi}
    F_i(\b{r})=u_0 \frac{N^3}{2} \left( f_1^i+i f_2^i\right) = u_0 \frac{N^3}{2} \sum_a \grad V_0(\b{r}-\b{R_a})\cdot \hat{q}_i e^{ i \b{q}_i \cdot \b{R}_a + i \phi}.
\end{align}
Now, using the symmetry of the system, we can establish the relation between $A_{ij}^n(\b{R})$, $n=1,2,3$. For monolayer, the point group symmetry is $D_{3h}$, with symmetry elements 
$\mathcal G \in \{E,C_3,\bar{C}_3, C_2, C_2', C_2'', S_3,\bar{S}_3, \sigma_v, \sigma_{v'}, \sigma_{v''}\}$.
\begin{align}
    A_{ij}^n(\b{R}) = \mel{\phi_i(\b{r})}{F_n(\b{r})}{\phi_j(\b{r}-\b{R})}
\end{align}
Under transformation operation $\mathcal G$:
\begin{align}
    A_{ij}^n(\mathcal G\b{R}) = \mel{\phi_i(\mathcal G\b{r})}{F_n(\mathcal G\b{r})}{\phi_j(g(\b{r}-\b{R}))} = \mel{\phi_k(\b{r}) \Gamma_{\mathcal G^{-1},ki}^{r_i}}{F_n(\mathcal G\b{r})}{\phi_l(\b{r}-\b{R}) \Gamma_{\mathcal G^{-1},lj}^{r_j}},
\end{align}
where $r_i, r_j$ stand for the representation of $\phi_i, \phi_j$ functions. The  transformation properties of $F_n(\b{r})$ are obtained using~\cref{eq: Fi} and are specfied in Table ~\ref{tab: Ftransform}. 
\begin{table}[!htbp]
\centering
\begin{tabular}{ |c|c|c|c|c|c|c|c|c|c|c|c| } 
 \hline
             & $C_3$ & $\bar{C}3$ & $C_2$ & $C_2'$ & $C_2''$ & $\sigma_h$ & $S_3$ & $\bar{S}_3$ & $\sigma_v$ & $\sigma_v'$ & $\sigma_v''$  \\ \hline
 $F_1(\b{r})$ & $F_3(C_3\b{r})$ & $F_2(\bar{C_3}\b{r})$  & $F_1(C_2\b{r})$  & $F_2(C_2'\b{r})$  &  $F_3(C_2''\b{r})$  &  $F_1(\b{r})$  &  $F_3(S_3\b{r})$  &  $F_2(\bar{S_3}\b{r})$ &  $F_1(\sigma_v \b{r})$  &  $F_2(\sigma_v' \b{r})$ &  $F_3(\sigma_v'' \b{r})$  \\ 
 $F_2(\b{r})$ & $F_1(C_3\b{r})$ & $F_3(\bar{C_3}\b{r})$  & $F_3(C_2\b{r})$  & $F_1(C_2'\b{r})$  &  $F_2(C_2''\b{r})$  &   --  &  --  &  -- &  $F_3(\sigma_v \b{r})$  &  $F_1(\sigma_v' \b{r})$ &  $F_2(\sigma_v'' \b{r})$ \\ 
 $F_3(\b{r})$ & $F_2(C_3\b{r})$ & $F_1(\bar{C_3}\b{r})$  & $F_2(C_2\b{r})$  & $F_3(C_2'\b{r})$  &  $F_1(C_2''\b{r})$  &   --  &  --  &  -- &  $F_2(\sigma_v \b{r})$  &  $F_3(\sigma_v' \b{r})$ &  $F_1(\sigma_v'' \b{r})$ \\ 
 \hline
\end{tabular}
\caption{Transformations of $F_i(\b{r})$, $i=1,2,3$.}
\label{tab: Ftransform}
\end{table}
Finally, in matrix form, we have
\begin{align}
    A^n(g\b{R}) = \Gamma_{g}^{r_i} A^{\tilde{n}}(R) \Gamma_{g}^{r_j \dagger},
\end{align}
where $\tilde{n}$ is found from~\cref{tab: Ftransform}. This allows us to symmetry-constrain the form of the CDW term:

For $\b{R}=0$, denoting $B^n=A^n(\b{0})$, $n=1,2,3$, we then find system of equations
\begin{align}
    B^1 & =\Gamma_{\sigma_\nu}^{r_i} B^1 \Gamma_{\sigma_\nu}^{r_j \dagger}, \ \ \ \ B^2  = \Gamma_{\sigma_\nu}^{r_i} B^3 \Gamma_{\sigma_\nu}^{r_j \dagger}, \\
    B^1 & =\Gamma_{\bar{C_3}}^{r_i} B^1 \Gamma_{\bar{C_3}}^{r_j \dagger},\ \ \ \ B^2  = \Gamma_{C_3}^{r_i} B^3 \Gamma_{C_3}^{r_j \dagger}, \\
    B^1 & =\Gamma_{\sigma_\nu'}^{r_i} B^2 \Gamma_{\sigma_\nu'}^{r_j \dagger}, \ \ \ \ B^2 =\Gamma_{\sigma_\nu''}^{r_i} B^2 \Gamma_{\sigma_\nu''}^{r_j \dagger}, \\
    B^1 & =\Gamma_{C_3}^{r_i} B^2 \Gamma_{C_3}^{r_j \dagger}, \ \ \ \ B^3  =\Gamma_{\sigma_\nu'}^{r_i} B^3 \Gamma_{\sigma_\nu'}^{r_j \dagger}, \\
    B^1 & =\Gamma_{\sigma_\nu''}^{r_i} B^3 \Gamma_{\sigma_\nu''}^{r_j \dagger}, \\
\end{align}

Solving this system, we find that
\begin{align}
\label{eq: B-matrix}
    B^1=B^2=B^3 = \begin{pmatrix}
        \tilde{b}_1 & 0 & 0 \\
        0 & \tilde{b}_2 & 0 \\
        0 & 0 & \tilde{b}_2
    \end{pmatrix},
\end{align}
where $\tilde{b}_1$ and $\tilde{b}_2$ are some complex numbers. Analogously, we can derive similar relations for vectors $\b{R}$ connecting nearest neighbours, next nearest neigbors, and so on. The free parameters like $\tilde{b}_1,\ \tilde{b}_2$ in~\cref{eq: B-matrix} then could be determined self-consistently in the BdG-like approach applied to the CDW Hamiltonian.

Overall, for the commensurate CDW, the Hamiltonian is
\begin{align}
    \hat{H} = \sum_\b{k} c_\b{k}^\dagger H_\b{k} c_\b{k} + \sum_{n,\b{k}} \left( c_{\b{k}+\b{q_n}}^\dagger \Lambda^n_\b{k} c_\b{k} + \mathrm{cc}.\right),
\end{align}
where $c_\b{k} = \left( c_{\b{k} z^2 \ua}, c_{\b{k} xy \ua}, c_{\b{k} x^2-y^2 \ua}, c_{\b{k} z^2 \da}, c_{\b{k} xy \da}, c_{\b{k} x^2-y^2 \da} \right)$, and $c^\dagger_{\b{k} \alpha s}$ creates an electron with momentum $\b{k}$ in orbital $\alpha$ and spin $s$.
In the space of vectors $\{ c_\b{k}, c_{\b{k}+\b{q_1}}, c_{\b{k}+\b{q_2}}, c_{\b{k}+\b{q_3}}, c_{\b{k}+\b{q_{12}}}, c_{\b{k}+\b{q_{13}}}, c_{\b{k}+\b{q_{23}}}, c_{\b{k}+\b{q_{\overbar{12}}}}, c_{\b{k}+\b{q_{\overbar{13}}}}\}$, the CDW Hamiltonian takes form 
\begin{align}
\label{eq: H_CDW}
    H_{\mathrm{CDW}}(\b{k}) = \begin{pmatrix}
        H_\b{k} & \Lambda^{1}_\b{k} & \Lambda^{2}_\b{k} & \Lambda^{3}_\b{k} & \Lambda^{3 \dagger}_{\b{k}+\b{q_{12}}} & \Lambda^{2 \dagger}_{\b{k}+\b{q_{13}}} & \Lambda^{1 \dagger}_{\b{k}+\b{q_{23}}} & 0 & 0 \\
        \Lambda^{1 \dagger}_\b{k} & H_{\b{k}+\b{q_1}} & 0 & 0 & \Lambda^{2}_{\b{k}+\b{q_1}} & \Lambda^{3}_{\b{k}+\b{q_1}} & \Lambda^{1}_{\b{k}+\b{q_1}} & \Lambda^{2 \dagger}_{\b{k}+\b{q_{\overbar{12}}}} & \Lambda^{3 \dagger}_{\b{k}+\b{q_{\overbar{13}}}} \\
        \Lambda^{2 \dagger}_\b{k} & 0 & H_{\b{k}+\b{q_2}} & 0 & \Lambda^{1}_{\b{k}+\b{q_2}} & \Lambda^{2}_{\b{k}+\b{q_2}} & \Lambda^{3}_{\b{k}+\b{q_2}} & \Lambda^{3 \dagger}_{\b{k}+\b{q_{\overbar{12}}}} & \Lambda^{1 \dagger}_{\b{k}+\b{q_{\overbar{13}}}} \\
        \Lambda^{3 \dagger}_\b{k} & 0 & 0 & H_{\b{k}+\b{q_3}} & \Lambda^{3}_{\b{k}+\b{q_3}} & \Lambda^{1}_{\b{k}+\b{q_3}} & \Lambda^{2}_{\b{k}+\b{q_3}} & \Lambda^{1 \dagger}_{\b{k}+\b{q_{\overbar{12}}}} & \Lambda^{2 \dagger}_{\b{k}+\b{q_{\overbar{13}}}} \\
        \Lambda^{3}_{\b{k}+\b{q_{12}}} & \Lambda^{2 \dagger}_{\b{k}+\b{q_1}} & \Lambda^{1 \dagger}_{\b{k}+\b{q_2}} & \Lambda^{3 \dagger}_{\b{k}+\b{q_3}} & H_{\b{k}+\b{q_{12}}} & 0 & 0 & \Lambda^{2}_{\b{k}+\b{q_{12}}} & \Lambda^{1}_{\b{k}+\b{q_{12}}} \\
        \Lambda^{2}_{\b{k}+\b{q_{13}}} & \Lambda^{3 \dagger}_{\b{k}+\b{q_1}} & \Lambda^{2 \dagger}_{\b{k}+\b{q_2}} & \Lambda^{1 \dagger}_{\b{k}+\b{q_3}} & 0 & H_{\b{k}+\b{q_{13}}} & 0 & \Lambda^{1}_{\b{k}+\b{q_{13}}} & \Lambda^{3}_{\b{k}+\b{q_{13}}} \\
       \Lambda^{1}_{\b{k}+\b{q_{23}}} & \Lambda^{1 \dagger}_{\b{k}+\b{q_1}} & \Lambda^{3 \dagger}_{\b{k}+\b{q_2}} & \Lambda^{2 \dagger}_{\b{k}+\b{q_3}} & 0 & 0 & H_{\b{k}+\b{q_{23}}} & \Lambda^{3}_{\b{k}+\b{q_{23}}} & \Lambda^{2}_{\b{k}+\b{q_{23}}} \\
        0 & \Lambda^{2}_{\b{k}+\b{q_{\overbar{12}}}} & \Lambda^{3}_{\b{k}+\b{q_{\overbar{12}}}} & \Lambda^{1}_{\b{k}+\b{q_{\overbar{12}}}} & \Lambda^{2 \dagger}_{\b{k}+\b{q_{12}}} & \Lambda^{1 \dagger}_{\b{k}+\b{q_{13}}} & \Lambda^{3 \dagger}_{\b{k}+\b{q_{23}}} & H_{\b{k}+\b{q_{\overbar{12}}}} & 0 \\
        0 & \Lambda^{3}_{\b{k}+\b{q_{\overbar{13}}}} & \Lambda^{1}_{\b{k}+\b{q_{\overbar{13}}}} & \Lambda^{2}_{\b{k}+\b{q_{\overbar{13}}}} & \Lambda^{1 \dagger}_{\b{k}+\b{q_{12}}} & \Lambda^{3 \dagger}_{\b{k}+\b{q_{13}}} & \Lambda^{2 \dagger}_{\b{k}+\b{q_{23}}} & 0 & H_{\b{k}+\b{q_{\overbar{13}}}}
    \end{pmatrix},
\end{align}
where $\b{k}$ is restricted to lBZ, $\b{k} \in \mathrm{lBZ}$, and we used notation $\b{q_{12}}=\b{q_1}+\b{q_2},\ \b{q_{13}}=\b{q_1}+\b{q_3},\ \b{q_{23}}=\b{q_2}+\b{q_3},\ \b{q_{\overbar{12}}}=\b{q_1}-\b{q_2},\ \b{q_{\overbar{13}}}=\b{q_1}-\b{q_3}$. 

\begin{figure*}[!htbp]
    \centering
    \begin{subfigure}[t]{0.49\textwidth}
    \centering
    \includegraphics[width=\linewidth]{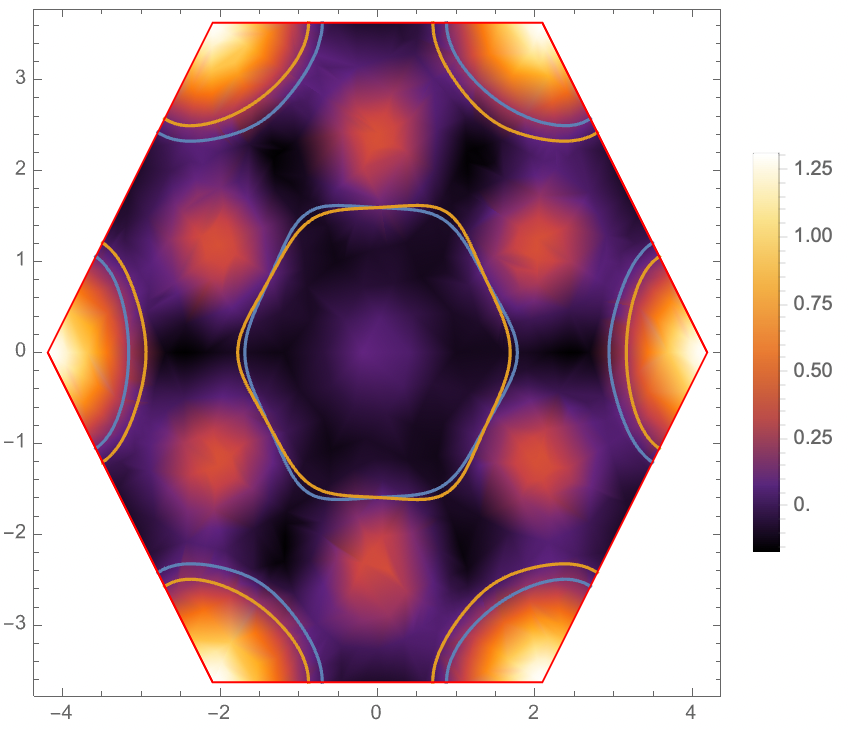}
    \caption{}
    \end{subfigure}
    \centering
    \begin{subfigure}[t]{0.49\textwidth}
    \centering
    \includegraphics[width=\linewidth]{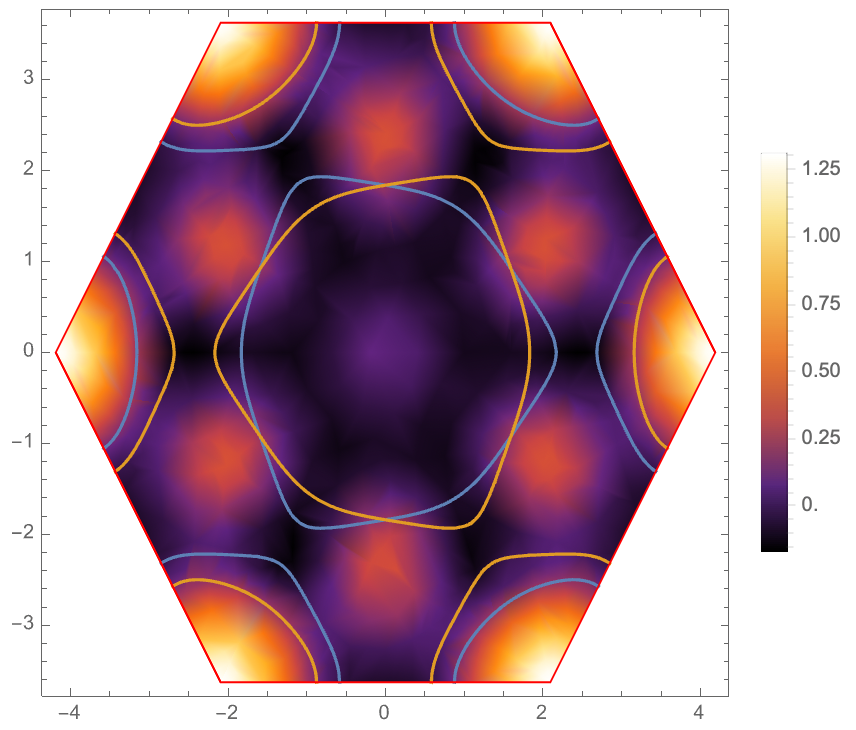}
    \caption{}
    \end{subfigure}
    \caption{Density plot of $\Lambda(\b{k})$ in 1BZ with superimposed FS for NbSe$_2$ (a) and TaS$_2$ (b). }
    \label{fig: Delta(k)}
\end{figure*}

The values of the CDW gaps $\Lambda^i_\b{k}$ can be determined self-consistently.  However, expecting that results of interest will not depend on the fine details of the CDW-term (an argument supporting this will be present in the next section), we use the functional form determined in~\cite{flicker2016charge} for NbSe$_2$ in the band basis  and adapt it to our approach. We set
\begin{align}
    \Lambda^1_\b{k} & = \Lambda^2_\b{k}=\Lambda^3_\b{k} = b \begin{pmatrix}
        \Lambda(\b{k}) & 0 & 0 \\
        0 &  \Lambda(\b{k}) & 0 \\
        0 & 0 & \Lambda(\b{k})
    \end{pmatrix},
\end{align}
where
\begin{align}
    \Lambda(\b{k}) & = c_1 (2 \cos (\alpha k_x) \cos (\beta k_y)+\cos (\alpha k_x))+c_2 (2 \cos (3 \alpha k_x) \cos (\beta k_y)+\cos (2 \beta k_y))+c_3 (2 \cos (2 \alpha k_x) \cos (2 \beta k_y)+\cos (4 \alpha k_x)) \\ \nn 
     &+c_4 (\cos (5 \alpha k_x) \cos (\beta k_y)+\cos (4 \alpha k_x) \cos (2 \beta k_y)+\cos (\alpha k_x) \cos (3 \beta k_y))+c_5 (2 \cos (3 \alpha k_x) \cos (3 \beta k_y)+\cos (6 \alpha k_x))+c_0, \\
     \alpha & =\frac{a}{2},\  \beta=\frac{\sqrt{3}a}{2}\,.
\end{align}
The parameter $b$ is considered as a phenomenological parameter hereafter. It is used to control the overall strength of the CDW-term. Following~\cite{flicker2016charge}, where the constants $c_i$ are determined self-consistently, we determine $c_i$ from the condition that $\Lambda(\b{k})$ is not zero only at one high-symmetry point, namely at $K$ we require $\Lambda(\b{k})=1.25$. We find 
\begin{align}
    c_0 & = 0.135;\  c_1  = -0.114; \\ \nn
    c_2 & = 0.116;\  c_3  = -0.060; \\ \nn
    c_4 & = -0.095;\  c_5  = 0.141;
\end{align}
We will use the same form of the CDW term for TaS$_2$ as well. We plot $\Lambda(\b{k})$ function with superimposed Fermi surfaces for NbSe$_2$ and TaS$_2$ in~\cref{fig: Delta(k)}.

\subsection{Superconductivity -- BCS theory}
Now we add interaction that leads to $s$-wave superconductivity. Using group theory arguments as outlined in Ref.~\cite{mockli2018robust}, one can conclude that allowed $s$-wave correlations (in real space) are
\begin{align}\expval{\Psi_{1,i}}=\expval{c_{iz^2,\ua} c_{iz^2,\da}},\  \mathrm{and} \  \expval{\Psi_{2,i}}=\expval{c_{ixy,\ua}c_{ixy,\da}}+\expval{c_{ix^2-y^2,\ua} c_{ix^2-y^2,\da}},
\end{align}
where $i$ is the index indicating the location in the lattice. We then can write the pairing contact interaction in the form
\begin{align}
\hat{H}_\mathrm{I}=-\sum_i \left[ U_1 \Psi_{1,i}^\dagger \Psi_{1,i} + U_2 \Psi_{2,i}^\dagger \Psi_{2,i} + U_3 \left( \Psi_{1,i}^\dagger \Psi_{2,i} + \Psi_{2,i}^\dagger \Psi_{1,i}  \right) \right]
\end{align}
Correspondingly, in the momentum space, restricting interaction to most relevant channel (i.e., the one corresponding to zero momentum of the center of mass of a pair), we write down effective interaction in the form

\begin{align}
    \hat{H}_\mathrm{I}=-\sum_{k,k'} \left[ U_1 \Psi_{1,\b{k}}^\dagger \Psi_{1,\b{k'}} + U_2 \Psi_{2,\b{k}}^\dagger \Psi_{2,\b{k'}} + U_3 \left( \Psi_{1,\b{k}}^\dagger \Psi_{2,\b{k'}} + \Psi_{2,\b{k}}^\dagger \Psi_{1,\b{k'}}  \right) \right],
\end{align}
where $\Psi_{1,\b{k}} = c_{z^2\b{k},\ua} c_{z^2-\b{k},\da}$ and $\Psi_{2,\b{k}}=c_{xy\b{k},\ua}c_{xy-\b{k},\da}+c_{x^2-y^2\b{k},\ua} c_{x^2-y^2-\b{k},\da}$.

Denoting $\Delta_i = U_i \expval{c_{i\b{p},\ua} c_{i-\b{p},\da}}$, where $i=1,2,3$ correspond to $z^2, xy, x^2-y^2$, respectively, in the mean-field treatment the interaction takes form

\begin{align}
    \hat{H}_{\mathrm{I,MF}} = \sum_\b{k} \left(\Delta_1+ \frac{U_3}{U_2} (\Delta_2+\Delta_3) \right) c_{1\b{k}\ua}^\dagger c_{1-\b{k}\da}^\dagger + \left(\Delta_2+\Delta_3+ \frac{U_3}{U_1} \Delta_1 \right) \left( c_{2\b{k}\ua}^\dagger c_{2-\b{k}\da}^\dagger +c_{3\b{k}\ua}^\dagger c_{3-\b{k}\da}^\dagger \right).
\end{align}

To proceed, we use the BdG formalism

\begin{align}
\label{eq: H_BDG}
    H_{\mathrm{BdG}}(\b{k}) = \frac{1}{2} \begin{pmatrix}
        H(\b{k}) & \hat{\Delta} \\
        \hat{\Delta}^\dagger & -H(-\b{k})^T
    \end{pmatrix},\ \  \mathrm{and} \ \  \hat{H} = \sum_{\b{k}} \Psi_\b{k}^\dagger H_{BdG}(\b{k}) \Psi_\b{k},
\end{align}
where $\hat{\Delta}$ is the pairing matrix and $\Psi_{\b{k}}$ is the fermion destruction operator in the Nambu space. In the case of no CDW,

\begin{align}
    \Psi_\b{k} & = \left( c_{1\b{k}\ua}, c_{2\b{k}\ua}, c_{3\b{k}\ua}, c_{1\b{k}\da}, c_{2\b{k}\da}, c_{3\b{k}\da}, c_{1-\b{k}\ua}^\dagger, c_{2-\b{k}\ua}^\dagger, c_{3-\b{k}\ua}^\dagger, c_{1-\b{k}\da}^\dagger, c_{2-\b{k}\da}^\dagger, c_{3-\b{k}\da}^\dagger  \right)^T, \\ \nn
    \hat{\Delta} & = i \sigma_y \otimes \Delta_o,\ \  \mathrm{where}\ \  \Delta_o = \begin{pmatrix}
        \Delta_1+ \frac{U_3}{U_2} (\Delta_2+\Delta_3) & 0 & 0 \\
        0 & \Delta_2+\Delta_3+ \frac{U_3}{U_1} \Delta_1 & 0  \\
        0 & 0 & \Delta_2+\Delta_3+ \frac{U_3}{U_1} \Delta_1 
    \end{pmatrix},
\end{align}
where $\sigma_y$ is the Pauli matrix.

When the CDW is present, the Nambu space is formed by 
\begin{align}
\Psi_\b{k}=\left( c_{\b{k_1}\ua}, ..., c_{\b{k_9}\ua}, c_{\b{k_1}\da}, ..., c_{\b{k_9}\da}, c_{-\b{k_1} \ua}^{\dagger T}, ... , c_{-\b{k_9} \ua}^{\dagger T}, c_{-\b{k_1} \da}^{\dagger T}, ... , c_{-\b{k_9} \da}^{\dagger T}  \right)^T,
\end{align}
where $c_\b{k_1s} = \left(c_\b{1k_1s}, c_\b{2k_1s}, c_\b{3k_1s}  \right)$ with $s$ being a spin index, $\b{k_1} \in \mathrm{lBZ}$ and $\b{k_2}, ..., \b{k_9}$ vectors belong to domains $\mathrm{lBZ}_2, ..., \mathrm{lBZ}_9$, respectively, in $\b{k}$-space formed by shifting $\mathrm{lBZ}$ by linear combinations of $\b{q_1},\b{q_2},\ \mathrm{and}\ \b{q_3}$, such that the union of these domains with $\mathrm{lBZ}$ covers the domain equivalent to $\mathrm{1BZ}$. We note that in this case $H(-\b{k})$ in~\cref{eq: H_CDW} is not simply formed by taking $\b{k} \ra -\b{k}$ in $H(\b{k})$ since, for example, $-\b{k}+\b{q_1} \neq -(\b{k}+\b{q_1})$: proper rearrangement of blocks is necessary.

In principle, one can work with such defined Hamiltonian numerically. But the size of the matrices is quite large, and practice showed that straightforward computations with this Hamiltonian defined in $k$-space takes too much time. However, the special case $U_1=U_2=U_3$ allows significant analytical advancement.

\subsubsection{Analytical solution in the case $U_1=U_2=U_3$}

Let us first rewrite~\cref{eq: H_BDG} more explicitly expanding in spin space: 

\begin{align}
    H_{\mathrm{BdG}}(\b{k}) = \frac{1}{2} \begin{pmatrix}
        H_{\ua}(\b{k}) & 0 & 0 & \hat{\Delta}_c \\
        0 & H_{\da}(\b{k}) & -\hat{\Delta}_c & 0 \\
        0 & -\hat{\Delta}_c^\dagger & -H_{\da}(\b{k}) & 0 \\
        \hat{\Delta}_c^\dagger & 0 & 0 & -H_{\ua}(\b{k})
    \end{pmatrix},
\end{align}
where $\Delta_c=I_9 \otimes \Delta_0$.
Here we also used the time-reversal symmetry relations to write down the bottom right block.
In the special case $U_1=U_2=U_3$, $\Delta_c$ becomes proportional to the identity matrix, $\Delta_c = \Delta \times I_{27}$, where $\Delta=\Delta_1+\Delta_2+\Delta_3$. In this case, in the band basis there is no inter-band pairings. Indeed, let 

\begin{align}
    U_\b{k}=\begin{pmatrix}
        U_{\b{k}\ua} & 0 \\ 
        0 & U_{\b{k}\da}
    \end{pmatrix}
\end{align}
be the unitary operator which rotates $H(\b{k})$ into the band basis $\b{a}_\b{k}=U \b{c}_\b{k}$. Then

\begin{align}
    \begin{pmatrix}
        0 & \hat{\Delta}_c \\
        -\hat{\Delta}_c & 0
    \end{pmatrix} \ra \begin{pmatrix}
        U_{\b{k}\ua} & 0 \\ 
        0 & U_{\b{k}\da}
    \end{pmatrix} \begin{pmatrix}
        0 & \hat{\Delta}_c \\
        -\hat{\Delta}_c & 0
    \end{pmatrix} \begin{pmatrix}
        U_{-\b{k}\ua}^T & 0 \\ 
        0 & U_{-\b{k}\da}^T 
    \end{pmatrix} = \begin{pmatrix}
        0 & U_{\b{k}\ua}\hat{\Delta}_cU_{\b{k}\ua}^\dagger \\
        -U_{\b{k}\da}\hat{\Delta}_c U_{\b{k}\da}^\dagger & 0
    \end{pmatrix} = \begin{pmatrix}
        0 & \hat{\Delta}_c \\
        -\hat{\Delta}_c & 0
    \end{pmatrix}.
\end{align}
In the band basis, the BdG Hamiltonian takes form
\begin{align}
\label{eq: H_BdG_band_basis}
    H_{\mathrm{BdG}}(\b{k}) = \frac{1}{2} \begin{pmatrix}
        \hat{\epsilon}_{\ua}(\b{k}) & 0 & 0 & \hat{\Delta}_c \\
        0 & \hat{\epsilon}_{\da}(\b{k}) & -\hat{\Delta}_c & 0 \\
        0 & -\hat{\Delta}_c^\dagger & -\hat{\epsilon}_{\da}(\b{k}) & 0 \\
        \hat{\Delta}_c^\dagger & 0 & 0 & -\hat{\epsilon}_{\ua}(\b{k})
    \end{pmatrix},
\end{align}
where $\hat{\epsilon}_{\ua/\da}(\b{k})$ is the diagonal matrix of eigenvalues $\epsilon_{i\ua/\da}(\b{k})$ corresponding to spin up/down. Permuting the basis, we can rewrite the BdG Hamiltonian in a more convenient way
\begin{align}
\label{eq: H_BdG_permuted}
    H_{\mathrm{BdG}}(\b{k}) = \frac{1}{2} \begin{pmatrix}
        \hat{\epsilon}_{\ua}(\b{k}) & \hat{\Delta}_c & 0 & 0 \\
        \hat{\Delta}_c^\dagger & -\hat{\epsilon}_{\ua}(\b{k}) & 0 & 0  \\
        0 & 0 & \hat{\epsilon}_{\da}(\b{k}) & -\hat{\Delta}_c  \\
        0 & 0 &  -\hat{\Delta}_c^\dagger & -\hat{\epsilon}_{\da}(\b{k}) 
    \end{pmatrix}
\end{align}
Squaring the Hamiltonian in~\cref{eq: H_BdG_permuted}, we can find the eigen-energies
\begin{align}
    E_{ns}(\b{k})^2=\epsilon_{ns}(\b{k})^2+\Delta^2, \ \ (n=1,...,27).
\end{align}
To find the eigenvectors, we have to solve

\begin{align}
    \begin{pmatrix}
        \hat{\epsilon}_{s}(\b{k}) & s \hat{\Delta}_c \\
        s \hat{\Delta}_c & -\hat{\epsilon}_{s}(\b{k})
    \end{pmatrix} \begin{pmatrix}
        \b{u} \\
        \b{v}
    \end{pmatrix} = E_n \begin{pmatrix}
        \b{u} \\
        \b{v}
    \end{pmatrix},
\end{align}
where $s=\pm$ correspond to spin $\ua,\da$. The solution is
\begin{align}
    u_i & = \frac{\epsilon_{ns}(\b{k})+E_n}{\left( \left( \epsilon_{ns}(\b{k}) + E_n \right)^2 + \Delta^2 \right)^{\frac{1}{2}}} \delta_{in}, \\ \nn
    v_i & = \frac{\Delta}{\left( \left( \epsilon_{ns}(\b{k}) + E_n \right)^2 + \Delta^2 \right)^{\frac{1}{2}}} \delta_{in}, \ \ (i=1,...,27).
\end{align}

We can write the transformation to the basis in which the BdG Hamiltonian~\cref{eq: H_BdG_permuted} is diagonal in the form

\begin{align}
\label{eq: Bogoliubov_transform}
    a_{\b{k}is} & =u_{\b{k}is}^{j\tau} \gamma_{\b{k} j \tau} + v_{\b{k}is}^{j\tau} \gamma_{-\b{k} j \tau}^\dagger, \\ \nn
    a_{-\b{k}is}^\dagger & =v_{-\b{k}is}^{j\tau*} \gamma_{\b{k} j \tau} + u_{-\b{k}is}^{j\tau*} \gamma_{-\b{k} j \tau}^\dagger,
\end{align}
with
\begin{align}
    u_{\b{k}is}^{j\tau} & = \frac{\epsilon_{is}(\b{k})+E_{is}}{\left( \left( \epsilon_{is}(\b{k}) + E_{is} \right)^2 + \Delta^2 \right)^{\frac{1}{2}}} \delta_{s\tau} \delta_{ij}, \\ \nn
    v_{\b{k}is}^{j\tau} & = \frac{\Delta}{\left( \left( \epsilon_{i\tau}(-\b{k}) + E_{i\tau} \right)^2 + \Delta^2 \right)^{\frac{1}{2}}} (1-2 \delta_{\da,\tau}) \delta_{s \bar{\tau}} \delta_{ij},
\end{align}
where $i,j = 1,...,27$ are band indices, and $s,\tau$ are spin indices.
In the next subsection we proceed calculating  $T_c$ in preparation for the final goal -- the ratio of ${H_{c2}}/{\Delta^2}$.

\subsubsection{Self-consistent determiniation of $T_c$}

The self-consistent equations from which $T_c$ is determined are

\begin{align}
    \Delta_\alpha = -U_\alpha \sum_{\b{k} \in \mathrm{BZ}} \expval{c_{\alpha-\b{k}\da} c_{\alpha\b{k}\ua}} = -U_\alpha \sum_i \sum_{\b{k_i} \in \mathrm{lBZ_i}} \expval{c_{\alpha-\b{k_i}\da} c_{\alpha\b{k_i}\ua}}, \ \ \alpha=1,2,3.
\end{align}

One can show that
\begin{align}
    \Delta = \sum_\alpha \Delta_\alpha = - U_1 \sum_{\b{k}\in \mathrm{lBZ}} \sum_{i=1}^{27} \expval{a_{-\b{k} i \da} a_{\b{k} i \ua}}.
\end{align}

Using~\cref{eq: Bogoliubov_transform}, we find

\begin{align}
    - U_1 \sum_{\b{k}\in \mathrm{lBZ}} \expval{a_{-\b{k} i \da} a_{\b{k} i \ua}} = U_1 \sum_{\b{k}} \expval{\sum_{j\tau} v_{\b{k}i\ua}^{j\tau} u_{-\b{k}i\da}^{j\tau} \gamma_{-\b{k} j \tau}^\dagger \gamma_{-\b{k} j \tau} + \sum_{j\tau} u_{\b{k}i\ua}^{j\tau} v_{-\b{k}i\da}^{j\tau} \gamma_{\b{k} j \tau} \gamma_{\b{k} j \tau}^\dagger}.
\end{align}
Using $\expval{\gamma_{\b{k} j \tau}^\dagger \gamma_{\b{k} j \tau}} = n_F(E_{j \tau}(\b{k}))$ and $\{ a_{\b{k} i s}, a_{\b{k'} j \tau}^\dagger\} = \delta_{ij} \delta_{s\tau} \delta_{\b{k} \b{k'}}$, we find

\begin{align}
    \Delta = U_1 \sum_{\b{k},ij\tau} v_{\b{k}i\ua}^{j\tau} u_{-\b{k}i\da}^{j\tau} \left( n_F(E_{j \bar{\tau}} (\b{k})) + n_F(E_{j \tau} (\b{k})) - 1\right).
\end{align}
And eventually

\begin{align}
    \Delta = U_1 \sum_{\b{k},i} \frac{\Delta}{2 E_{i\da}(\b{k})} \left( 1- n_F(E_{i \ua} (\b{k})) - n_F(E_{i \da} (\b{k}) \right).
\end{align}

At $T_c$, this equation reduces to

\begin{align}
\label{eq: Tc}
    \frac{1}{U_1} = \sum_{\b{k},i} \frac{1}{2 \abs{\epsilon_{i\da}(\b{k})}} \left( 1- n_F(\abs{\epsilon_{i \ua} (\b{k})}) - n_F(\abs{\epsilon_{i \da} (\b{k})}) \right).
\end{align}

Note that energies $\epsilon_{is}$ are measured with respect to the chemical potential.

\begin{figure*}[!htbp]
    \centering
    \begin{subfigure}[t]{0.49\textwidth}
    \centering
    \includegraphics[width=\linewidth]{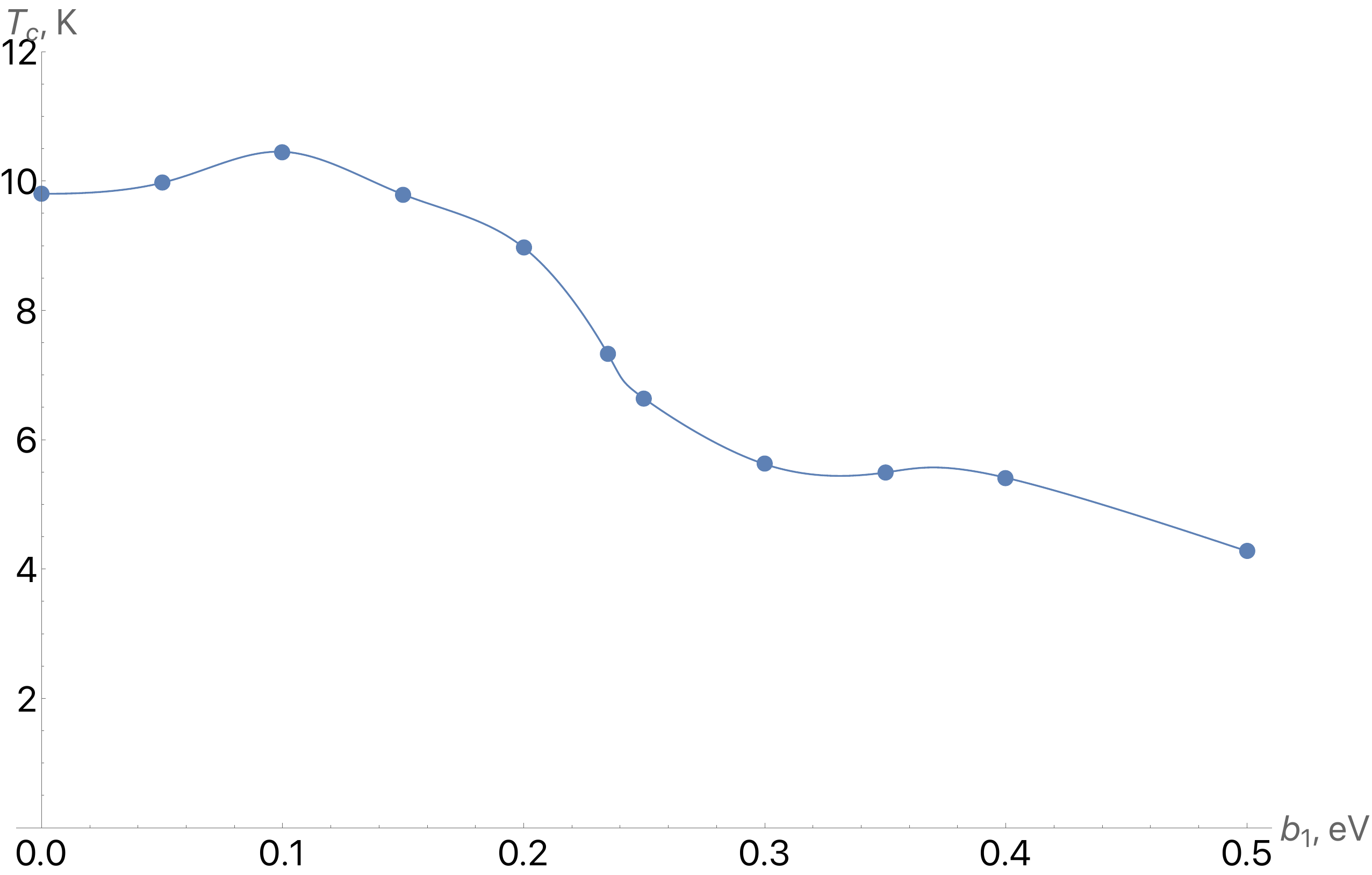}
    \caption{}
    \end{subfigure}
    \centering
    \begin{subfigure}[t]{0.49\textwidth}
    \centering
    \includegraphics[width=\linewidth]{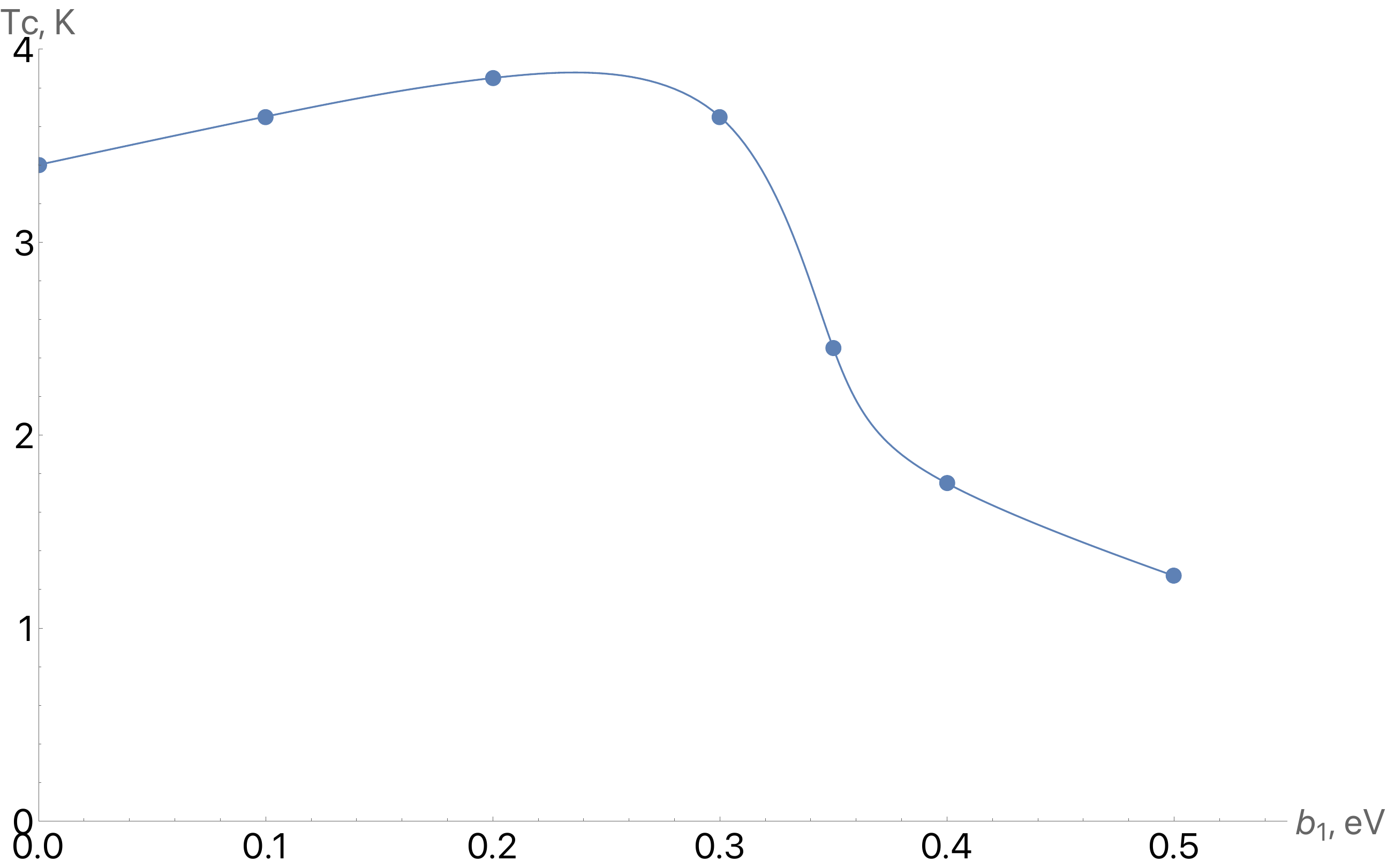}
    \caption{}
    \end{subfigure}
    \caption{$T_c$ as a function of CDW strength $b_1$ for (a) NbSe$_2$ and (b) TaS$_2$.}
    \label{fig: Tcvsb1}
\end{figure*}

We compute the dependence of $T_c$ on the CDW strength $b_1$ numerically. These dependencies are plotted in~\cref{fig: Tcvsb1}.

\begin{figure*}[!htbp]
    \centering
    \begin{subfigure}[t]{0.49\textwidth}
    \centering
    \includegraphics[width=\linewidth]{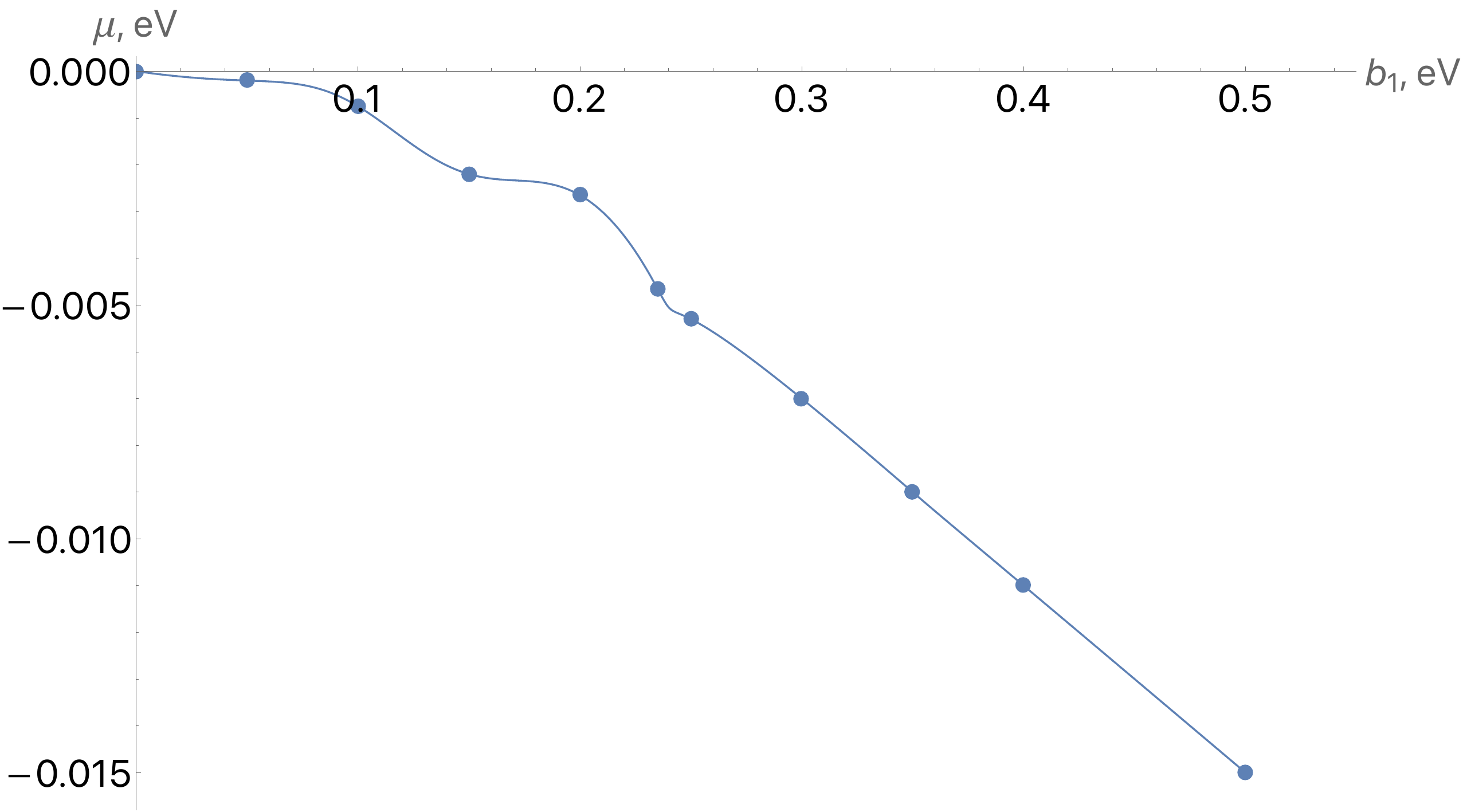}
    \caption{}
    \end{subfigure}
    \centering
    \begin{subfigure}[t]{0.49\textwidth}
    \centering
    \includegraphics[width=\linewidth]{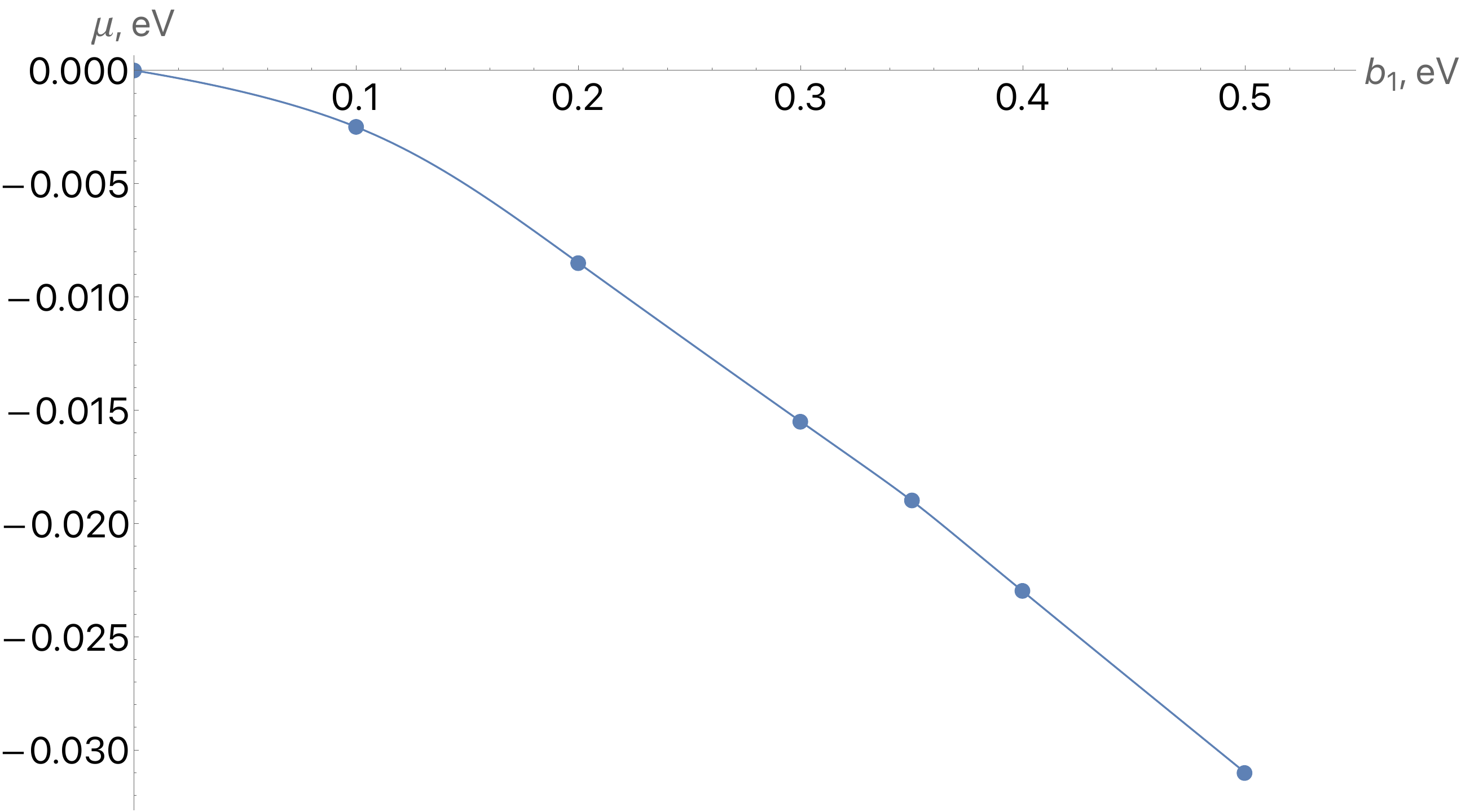}
    \caption{}
    \end{subfigure}
    \caption{$\mu$ as a function of CDW strength $b_1$ for (a) NbSe$_2$ and (b) TaS$_2$.}
    \label{fig: muvsb1}
\end{figure*}

\begin{figure*}[!htbp]
    \centering
    \begin{subfigure}[t]{0.49\textwidth}
    \centering
    \includegraphics[width=\linewidth]{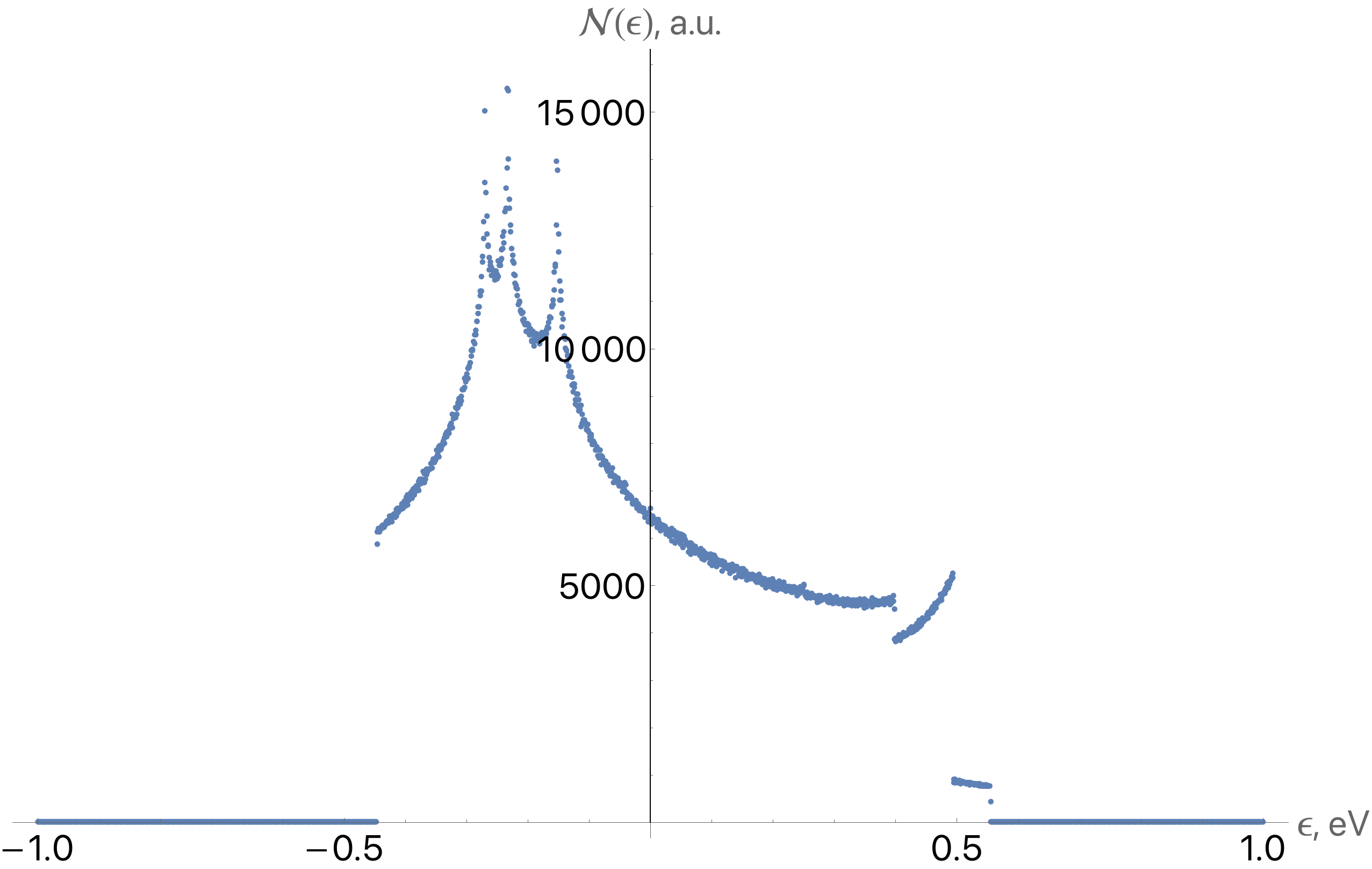}
    \caption{}
    \end{subfigure}
    \centering
    \begin{subfigure}[t]{0.49\textwidth}
    \centering
    \includegraphics[width=\linewidth]{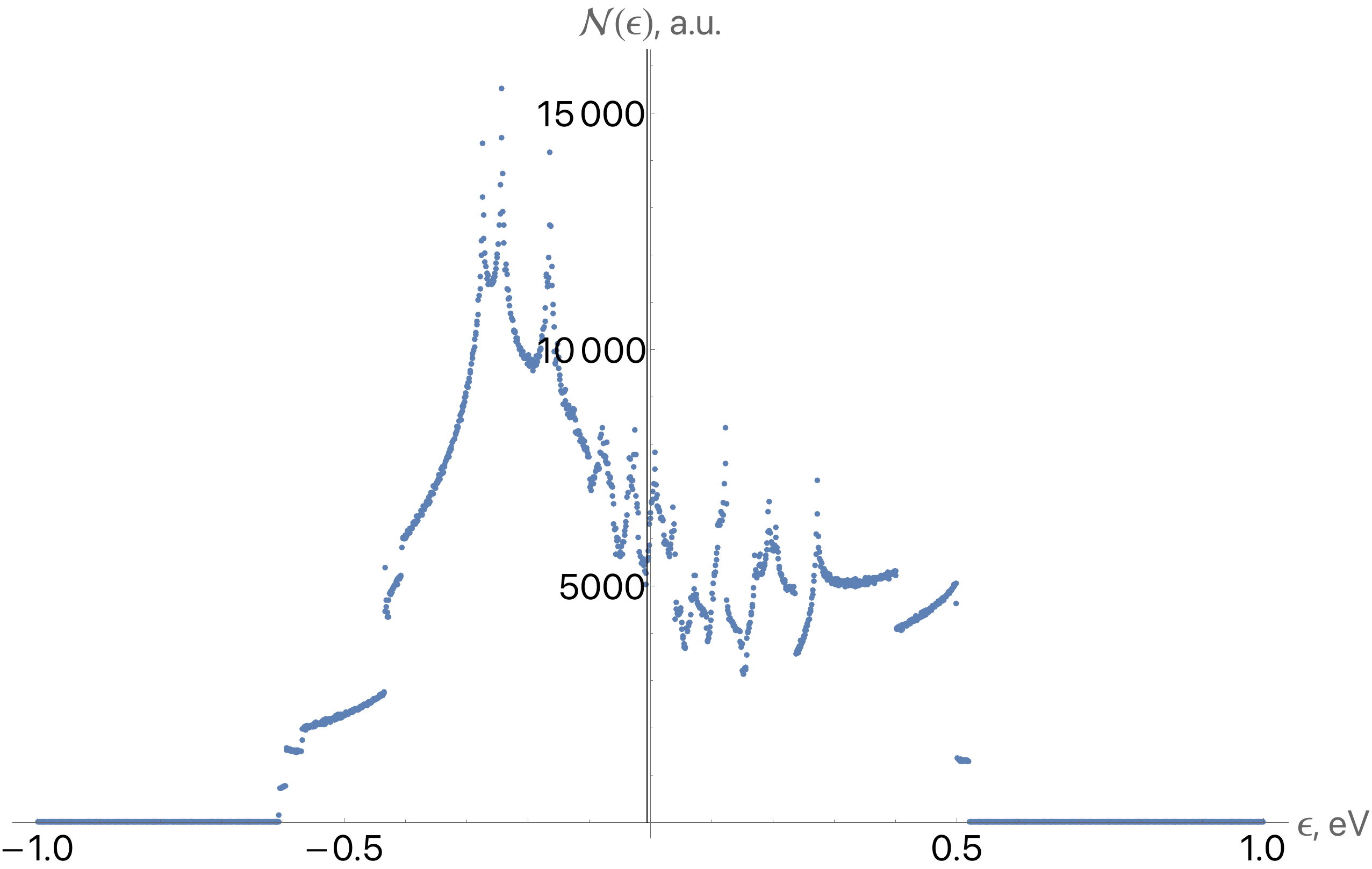}
    \caption{}
    \end{subfigure}
    \caption{DOS for NbSe$_2$ for CDW strength (a) $b_1=0$ eV and (b) $b_1=0.25$ eV.}
    \label{fig: DOS_NbSe2}
\end{figure*}

\begin{figure*}[!htbp]
    \centering
    \begin{subfigure}[t]{0.49\textwidth}
    \centering
    \includegraphics[width=\linewidth]{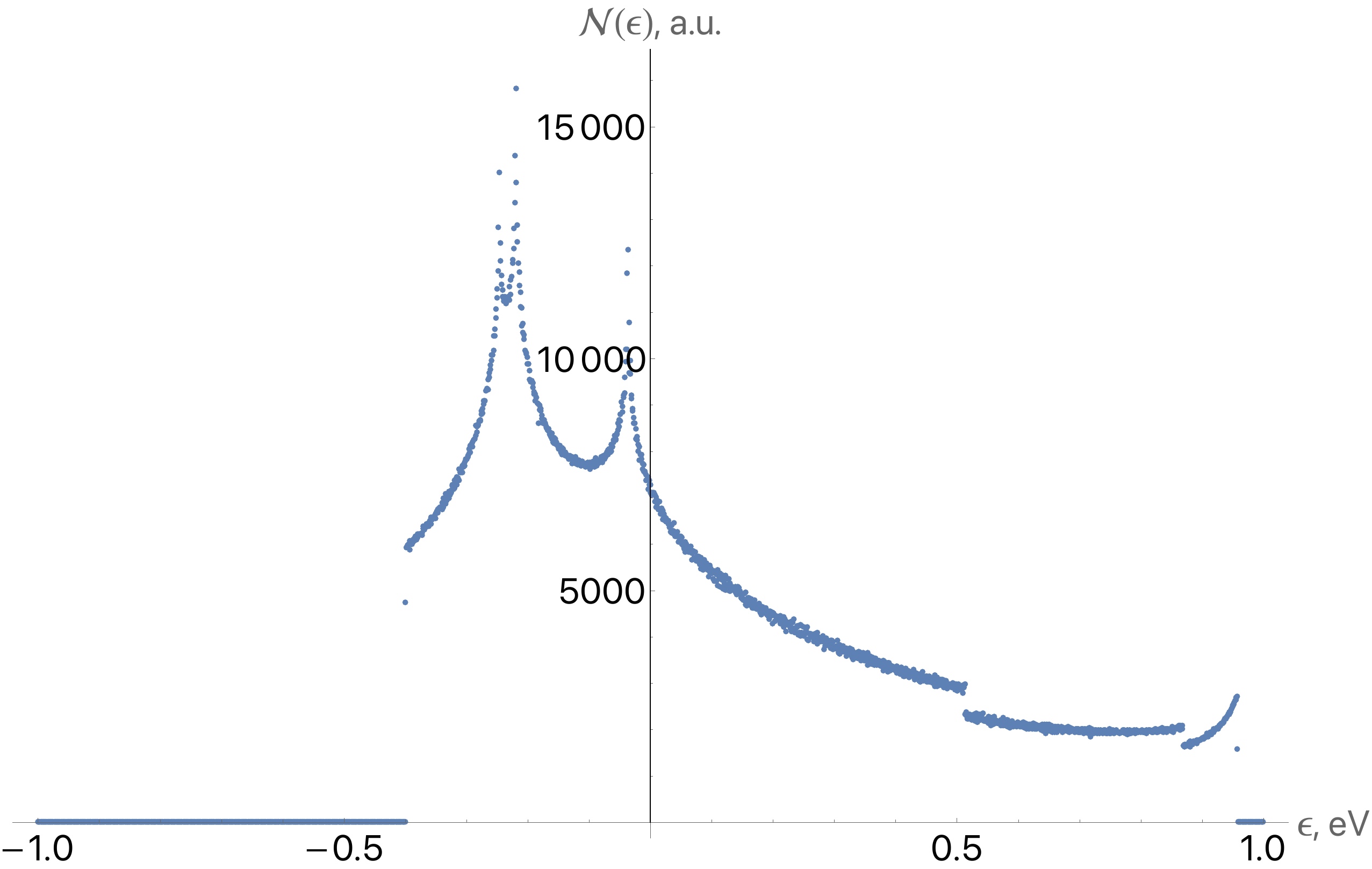}
    \caption{}
    \end{subfigure}
    \centering
    \begin{subfigure}[t]{0.49\textwidth}
    \centering
    \includegraphics[width=\linewidth]{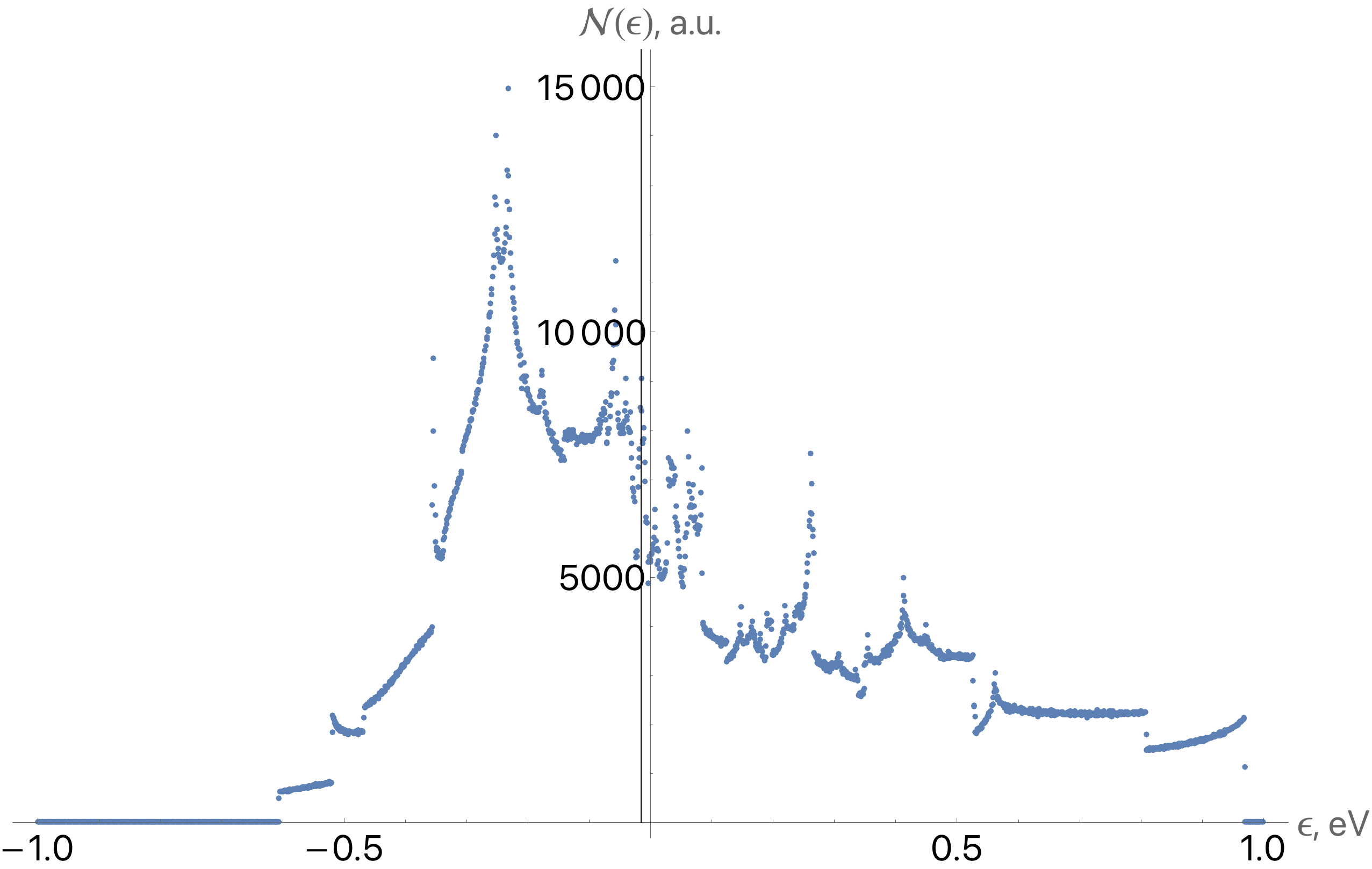}
    \caption{}
    \end{subfigure}
    \caption{DOS for TaS$_2$ for CDW strength (a) $b_1=0$ eV and (b) $b_1=0.3$ eV. 
    }
    \label{fig: DOS_TaS2}
\end{figure*}

\begin{figure*}[!htbp]
    \centering
    \begin{subfigure}[t]{0.49\textwidth}
    \centering
    \includegraphics[width=\linewidth]{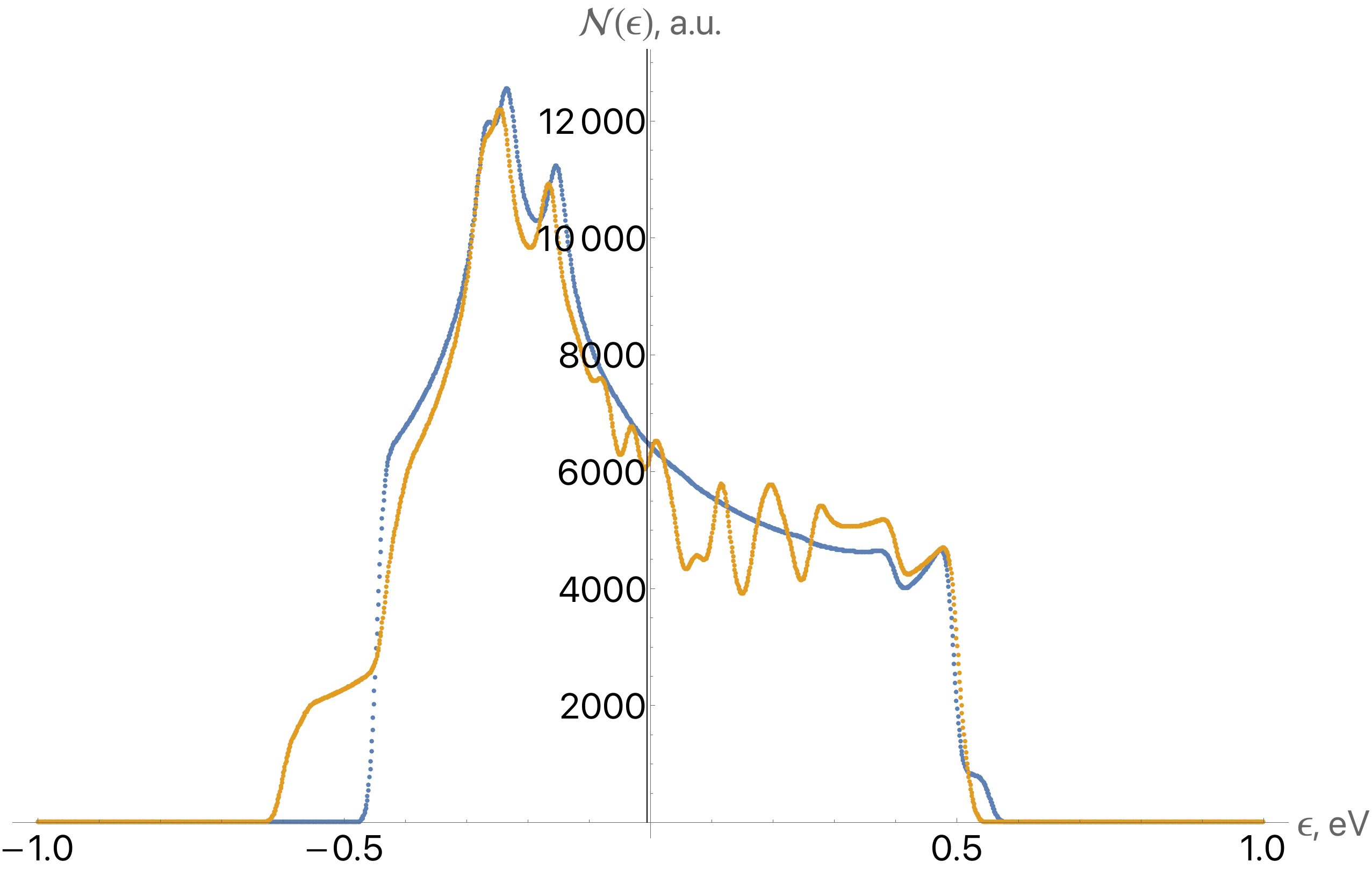}
    \caption{}
    \end{subfigure}
    \centering
    \begin{subfigure}[t]{0.49\textwidth}
    \centering
    \includegraphics[width=\linewidth]{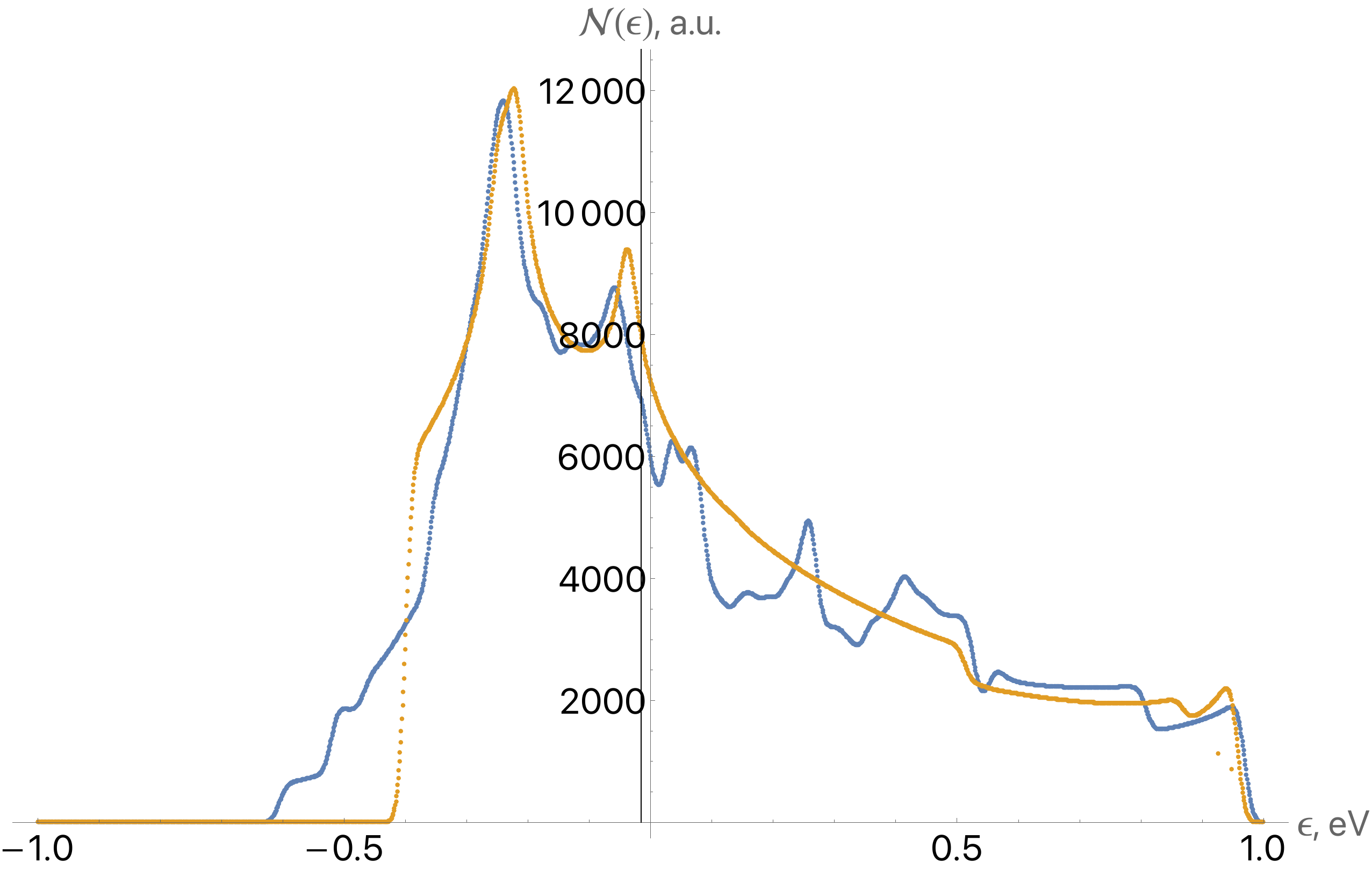}
    \caption{}
    \end{subfigure}
    \caption{Smeared DOS for (a) NbSe$_2$ at $b_1=0$ eV (blue) and $b_1=0.25$ eV (yellow) and for (b) TaS$_2$ at $b_1=0$ eV (yellow) and $b_1=0.3$ eV (blue). 
    }
    \label{fig: DOS_smeared}
\end{figure*}

Mainly, for a fixed value of $U_1$ the value of $T_c$ is controlled by the density of states $\mathcal{N}(\epsilon_F)$. The increase of $T_c$ for small values of $b_1$ happens due to the chemical potential, $\mu$, falls onto the van Hove singularity developing due to CDW order. The dependence of $\mu$ on the CDW strength $b_1$ is plotted in~\cref{fig: muvsb1}, and the density of states for some values of $b_1$ are plotted in~\cref{fig: DOS_NbSe2,fig: DOS_TaS2} for NbSe$_2$ and TaS$_2$, respectively. In~\cref{fig: DOS_smeared}, we plot the smeared DOS for NbSe$_2$ and TaS$_2$.

\subsection{Superconductivity -- Ginzburg-Landau theory}

Our goal is to compute the ratio of $H_{c2}/\Delta_0^2$ at the temperature at which the experimental measurements were conducted (which is essentially $T=0$). However, we will do this only approximately: Our line of attack is to compute $H_{c2}/\Delta^2$ at $T_c$ from the Ginzburg-Landau theory
\[ H_{c2}(T) \sim {\phi_0 \over 2\pi \xi_{GL}^2}(1-T/T_c)\,\,,\, \Delta(T) \sim \Delta_0 \sqrt{1-T/T_c} \; ,\;T\to T_c \]
and extrapolate to zero temperature. Namely, estimate the above equation at $T=0$, $H_{c2}(0)$. In actuality, this is an upper bound on this ratio, as it curves and saturates to a smaller value.

The form of the GL theory, which is consistent with the point group of the system, is given by 
\begin{align}
\label{eq: GL}
    S=\int d^dx d\tau \left( \frac{1}{2\eta} \abs{\left(-i \grad -2e A\right) \Delta}^2 + r(T) \abs{\Delta}^2 + \frac{u(T)}{2} \abs{\Delta}^4 + ...  \right)
\end{align}
Having $H_{c2} = {\phi_0}/{2 \pi \xi^2}$, $\Delta^2 = -{r(T)}/{u(T)}$, $\xi^2=-{1}/{2 \eta r(T)} = \xi_{GL}^2 (1-T/T_c)$, we find that the ratio of the upper-critical field to $\Delta^2$, in terms of the GL parameters, is given by
\begin{align}
    \frac{H_{c2}}{\Delta^2} = \frac{\phi_0}{\pi} \eta u.
\end{align}
Thus, our next (and final) goal is to compute these two coefficients from the microscopic model. 

\subsubsection{Quadratic terms:}
The coefficient of the quadratic terms in~\cref{eq: GL} are obtained from
\begin{align}
\label{eq: (G0Delta)^2}
    \Tr\left( G_0 \hat{\Delta} \right)^2 = -2 \Tr\left( G_{e\ua\b{p}} \hat{\Delta}_c G_{h\da\b{p-k}} \hat{\Delta}_c + G_{e\da\b{p}} \hat{\Delta}_c G_{h\ua\b{p-k}} \hat{\Delta}_c\right),
\end{align}

where
\begin{align}
    G_{e s \b{p}} = \left( -i \omega + H_{\mathrm{CDW},s}(\b{p}) \right)^{-1},\ G_{h s \b{p}} = \left( -i \omega - H^T_{\mathrm{CDW},s}(-\b{p}) \right)^{-1},
\end{align}
where $s=\ua,\da$, and $G_0$ is the normal part of the Green's function of the BDG Hamiltonian
\begin{align}
    G_0 = 2 G_e \oplus G_h,
\end{align}
where
\begin{align}
    G_e = G_{e \ua} \oplus G_{e \da},\ G_h = G_{h \ua} \oplus G_{h \da}.
\end{align}

Recall that for $U_1=U_2=U_3$, $\hat{\Delta}_c = I_{27} \otimes \Delta$. Then~\cref{eq: (G0Delta)^2} simplifies to 

\begin{align}
    \Tr\left( G_0 \hat{\Delta} \right)^2 = -2 \Tr\left( G_{e\ua\b{p}} G_{h\da\b{p-k}} + G_{e\da\b{p}} G_{h\ua\b{p-k}} \right) \Delta^2.
\end{align}

Denoting $\Pi_{s}(\b{k}) = \Tr\left( G_{es\b{p}} G_{h\bar{s}\b{p-k}}  \right)$ and switching to the band-basis (recall that $\hat{\Delta}_c$ is invariant under this transformation in the particular case we consider), we can write

\begin{align}
\label{eq: Ps(k)}
    \Pi_{s}(\b{k}) = \frac{T}{L^3} \sum_{\omega_n} \sum_{\b{p}} \sum_i \frac{1}{-i \omega_n + \epsilon_{i\b{p}s}} \frac{1}{-i \omega_n - \epsilon_{i\b{p}-\b{k}s}} \approx \frac{T}{L^3} \sum_{\omega_n} \sum_{\b{p}} \sum_i \frac{1}{-i \omega_n + \epsilon_{i\b{p}s}} \frac{1}{-i \omega_n - \epsilon_{i\b{p}s}+\b{v}_i(\b{p}) \cdot \b{k}}, 
\end{align}
and
\begin{align}
\label{eq: eta}
    \frac{1}{\eta} = - \left( \frac{\partial^2 \Pi_{\ua}(\b{0})}{\partial k_x^2} + \frac{\partial^2 \Pi_{\da}(\b{0})}{\partial k_x^2} \right).
\end{align}

Expanding in small $\b{k}$ and switching from the sum to the integration over the momentum

\begin{align}
    \frac{1}{2} \frac{\partial^2 \Pi_{s}(\b{0})}{\partial k_x^2} \approx \frac{1}{2} T \sum_i \sum_{\omega_n} \int p dp d\theta   \frac{1}{-i \omega_n + \epsilon_{i\b{p}s}} \frac{ v_i(p,\theta)^2 \cos^2\alpha_i(\theta) }{\left( -i \omega_n - \epsilon_{i\b{p}s} \right)^3},
\end{align}
where $\alpha_i(\theta)$ is the angle between $\b{v}_i$ and $\b{k}$.

One can show 
that $\frac{1}{2} \frac{\partial^2 \Pi_{s}(\b{k})}{\partial k_x^2} = \frac{1}{2} \frac{\partial^2 \Pi_{s}(\b{k})}{\partial k_y^2}$, and then we can write

\begin{align}
    \frac{\partial^2 \Pi_{s}(\b{0})}{\partial k_x^2} = \frac{1}{2} \frac{\partial^2 \Pi_{s}(\b{0})}{\partial k_x^2} + \frac{1}{2} \frac{\partial^2 \Pi_{s}(\b{0})}{\partial k_y^2} =  \frac{1}{2} T \sum_i \sum_{\omega_n} \int p dp d\theta   \frac{1}{-i \omega_n + \epsilon_{i\b{p}s}} \frac{ v_i(p,\theta)^2 }{\left( -i \omega_n - \epsilon_{i\b{p}s} \right)^3}.
\end{align}
Assuming $v_i(p,\theta)$ does not depend much on the absolute value of the momentum $\b{p}$ such that $v_i(p,\theta) \approx v_i(\theta)$ is a good approximation, we can rewrite the last expression using $\theta$-dependent density of states per angle introduced as follows

\begin{align}
    \int d^2 \b{k} f(\epsilon_\b{k},\theta) & = \int kdk \int d \theta f(\epsilon_\b{k},\theta) = \int kdk \int d \theta \int d\epsilon \delta(\epsilon-\epsilon_k) f(\epsilon,\theta) \\ \nn
    & =  \int d\epsilon \int d \theta \int kdk \ \delta(\epsilon-\epsilon_k) f(\epsilon,\theta) = \int d\epsilon \int d \theta \mathcal{N}_{\theta}(\epsilon) f(\epsilon,\theta),
\end{align}
where $f(\epsilon_\b{k},\theta)$ is an arbitrary function. The approximation $v_i(p,\theta) \approx v_i(\theta)$ is not valid in the vicinity of a Van Hove singularity.

Thus
\begin{align}
\label{eq: eta_approx}
    \frac{\partial^2 \Pi_{s}(\b{0})}{\partial k_x^2} & = \frac{1}{2} T \sum_i \sum_{\omega_n} \int_{-\omega_D}^{\omega_D} d \epsilon_i \int d\theta   \frac{\mathcal{N}_{i s \theta} (\epsilon_i)}{-i \omega_n + \epsilon_i} \frac{ v_i(\theta)^2 }{\left( -i \omega_n - \epsilon_i \right)^3}   \approx   \frac{1}{2} T \sum_i \sum_{\omega_n} \int_{-\omega_D}^{\omega_D} d \epsilon_i \int d\theta   \frac{\mathcal{N}_{i s \theta} (0)}{-i \omega_n + \epsilon_i} \frac{ v_i(\theta)^2 }{\left( -i \omega_n - \epsilon_i \right)^3} \\ \nn
    & = \frac{1}{2} T \sum_i \sum_{\omega_n} \int_{-\omega_D}^{\omega_D} d \epsilon_i   \frac{\mathcal{N}_{i s} (0)}{-i \omega_n + \epsilon_i} \frac{ \overbar{v_i^2} }{\left( -i \omega_n - \epsilon_i \right)^3} = \frac{1}{2} T \sum_{\omega_n} \int_{-\omega_D}^{\omega_D} d \epsilon   \frac{\mathcal{N}_{s} (0)}{-i \omega_n + \epsilon} \frac{ \overbar{v^2} }{\left( -i \omega_n - \epsilon \right)^3},
\end{align}
where $\mathcal{N}_{i s \theta}$ is the DOS per angle associated with the band $i$ and spin $s$, and we introduced average velocities associated with band $i$ via $\int d \theta \mathcal{N}_{is \theta} (0) v_i(\theta)^2 = \mathcal{N}_{is} (0) \overbar{v_i^2}$ and overall average velocity $\sum_i \mathcal{N}_{is} (0) \overbar{v_i^2} = \mathcal{N}_{s} (0) \overbar{v^2}$. We can proceed calculating the integral and eventually obtaining the Gorkov's result, but $\overbar{v^2}$ has to be computed numerically.

\cref{eq: eta_approx} can be used for gaining some insights into the effect of CDW if Van Hove singularities are sufficiently far away from the chemical potential level. In any case, we compute $\eta$ via numerically differentiating~\cref{eq: Ps(k)}.

\subsubsection{Quartic term:}

The quartic term is obtained from the static uniform part, $S$, of

\begin{align}
\label{eq: (G0Delta)^4}
    \Tr\left( G_0 \hat{\Delta} \right)^4 & = \Tr \bigg\{ G_{e\b{p}} \hat{\Delta}(\b{k_1}) G_{h\b{p}-\b{k_1}} \hat{\Delta}^\dagger(k_2)) G_{e\b{p}-\b{k_1}-\b{k_2}} \hat{\Delta}(\b{k_3}) G_{h\b{p}-\b{k_1}-\b{k_2}-\b{k_3}} \hat{\Delta}^\dagger(k_4)) \\ \nn
    & + G_{h\b{p}} \hat{\Delta}(\b{k_1}) G_{e\b{p}-\b{k_1}} \hat{\Delta}^\dagger(k_2)) G_{h\b{p}-\b{k_1}-\b{k_2}} \hat{\Delta}(\b{k_3}) G_{e\b{p}-\b{k_1}-\b{k_2}-\b{k_3}} \hat{\Delta}^\dagger(k_4)) \bigg\}.
\end{align}

Using band basis, we find

\begin{align}
    S = 2 \Tr\left\{ G_{e\ua\b{p}} G_{h\da\b{p}} G_{e\ua\b{p}} G_{h\da\b{p}} + G_{e\da\b{p}} G_{h\ua\b{p}} G_{e\da\b{p}} G_{h\ua\b{p}} \right\} \abs{\Delta}^4.
\end{align}

Denoting $\Sigma_s = \Tr\left\{ G_{es\b{p}} G_{h\bar{s}\b{p}} G_{es\b{p}} G_{h\bar{s}\b{p}} \right\} $, we have

\begin{align}
    u=4 (\Sigma_\ua + \Sigma_\da) = 8 \Sigma_\ua,
\end{align}
where the last equality holds due to TRS.

\begin{figure*}[!htbp]
    \centering
    \begin{subfigure}[t]{0.49\textwidth}
    \centering
    \includegraphics[width=\linewidth]{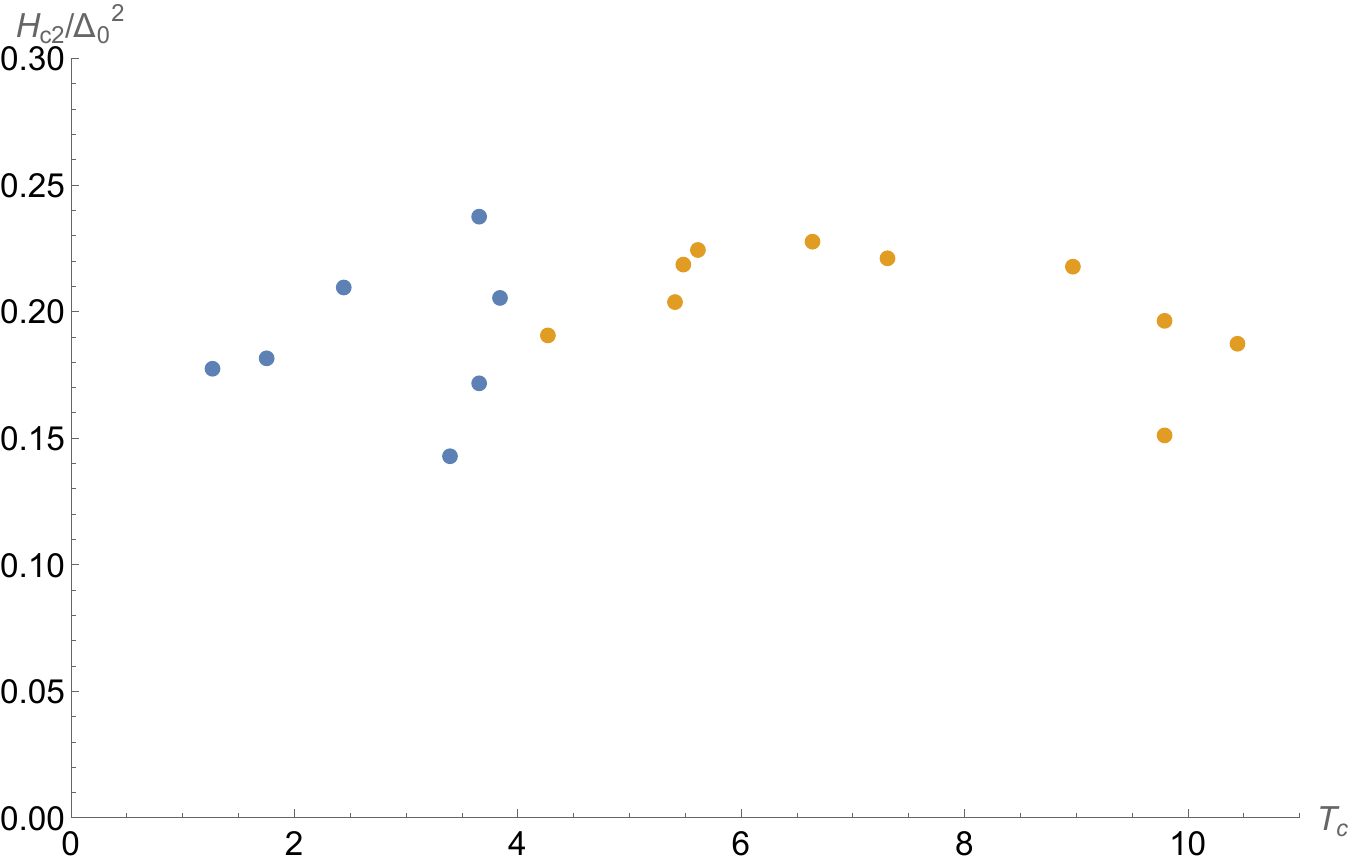}
    \end{subfigure}
    \caption{$H_{c2}/\Delta^2$ vs $T_c$ for TaS$_2$ (blue) and NbSe$_2$ (orange).}
    \label{fig: Hc2toDelta2}
\end{figure*}

In the band basis
\begin{align}
\label{eq: Sigma_s}
    \Sigma_s & = \frac{T}{L^3} \sum_i \sum_{\omega_n} \sum_{\b{p}} \frac{1}{\left( -i \omega_n + \epsilon_{i\b{p}s}\right)^2} \frac{1}{\left( -i \omega_n - \epsilon_{i\b{p}s}\right)^2} \\ \nn
    & = \frac{T}{L^3} \sum_i \sum_{\omega_n} \int_{-\omega_D}^{\omega_D} d\epsilon_{is} \frac{\mathcal{N}_{is}(\epsilon_{i s})}{\left( -i \omega_n + \epsilon_{i s}\right)^2} \frac{1}{\left( -i \omega_n - \epsilon_{i s}\right)^2} = \frac{T}{L^3} \sum_{\omega_n} \int_{-\omega_D}^{\omega_D} d\epsilon \frac{\mathcal{N}_{s}(\epsilon)}{\left( -i \omega_n + \epsilon\right)^2} \frac{1}{\left( -i \omega_n - \epsilon \right)^2} \\ \nn 
    & \approx \frac{T}{L^3} \sum_{\omega_n} \int_{-\omega_D}^{\omega_D} d\epsilon \frac{\mathcal{N}_{s}(0)}{\left( -i \omega_n + \epsilon\right)^2} \frac{1}{\left( -i \omega_n - \epsilon \right)^2}.
\end{align}
Completing this integral and obtaining  the Gor'kov's result,  we compute $u$ numerically from the first line of equation~\cref{eq: Sigma_s}.
Finally, we plot the computed $H_{c2}/\Delta^2$  for NbSe$_2$ and TaS$_2$ in~\cref{fig: Hc2toDelta2}.

\subsection{Details of computations}

In numerical computations, we utilize $C_3$ symmetry of the problem and generate Monkhorst-Pack grid~\cite{monkhorst1976special} for the one third of the lBZ with $L(L+1)$ points. For computing of all values, we used $L=600$.

The DOS were calculated by counting number of states $N_E$ in the intervals of energy $(E, E+\Delta E)$ with $\Delta E = 0.001$ eV from $E_0=-1$ eV to $E_f=1$ eV. The smeared DOS were calculated from the computed ones by smearing the delta functions in $N_E = N_E \int \delta \left(\epsilon-(E+\frac{\Delta E}{2}) \right) d\epsilon$ into normal distribution functions with dispersion $\sigma = 0.01$ eV.

The shift of the chemical potential for non-zero CDW strength was calculated by preserving number of occupied states on the same grid with $L=600$.

In computing critical temperatures,  
we set $T_c = 9.8\ (3.4)$ K for NbSe$_2$ (TaS$_2$) for zero CDW strength, $b_1=0$, and compute the corresponding $U_1$ from~\cref{eq: Tc}. For a non-zero CDW strength, $b_1 \neq 0$, we then use the computed $U_1$ to find $T_c$ from~\cref{eq: Tc}.

In computing $\eta$ from~\cref{eq: eta}, we use the central difference formula for the second derivative with $\Delta k_x = 10^{-9} \frac{3\sqrt{3}}{4} v_1$, where $v_1$ is the length of the primitive vector in the reciprocal lattice that defines lBZ.

\bibliography{bibliography}